\documentclass[%
 aip,
 amsmath,amssymb,
 reprint,%
]{revtex4-1}

\usepackage{graphicx}
\usepackage{dcolumn}
\usepackage{bm}

\usepackage[utf8]{inputenc}
\usepackage[T1]{fontenc}
\usepackage{mathptmx}
\usepackage{etoolbox}
\usepackage{xcolor}
\newcommand{\hip}{\texttt{HIPPIE} }

\makeatletter
\def\@email#1#2{%
 \endgroup
 \patchcmd{\titleblock@produce}
  {\frontmatter@RRAPformat}
  {\frontmatter@RRAPformat{\produce@RRAP{*#1\href{mailto:#2}{#2}}}\frontmatter@RRAPformat}
  {}{}
}%
\makeatother
\begin{document}

\preprint{AIP/123-QED}

\title[Electron temperature measurements in magnetic reconnection experiments at the NIF]{X-ray imaging and electron temperature evolution in laser-driven magnetic reconnection experiments at the National Ignition Facility}
\author{Vicente Valenzuela-Villaseca*}
 \affiliation{ 
Department of Astrophysical Sciences, Princeton University, Princeton, New Jersey 08544, USA.
}%
 \email{v.valenzuela@princeton.edu.}
 
\author{Jacob M. Molina}%
 \affiliation{ 
Department of Astrophysical Sciences, Princeton University, Princeton, New Jersey 08544, USA.}%
 \affiliation{ 
Princeton Plasma Physics Laboratory, Princeton, New Jersey 08540, USA.}

\author{Derek B. Schaeffer}%
 \affiliation{Department of Physics and Astronomy, University of California Los Angeles, Los Angeles, California 90095, USA.}

\author{Sophia Malko}%
 \affiliation{ 
Princeton Plasma Physics Laboratory, Princeton, New Jersey 08540, USA.}

\author{Jesse Griff-McMahon}%
 \affiliation{ 
Department of Astrophysical Sciences, Princeton University, Princeton, New Jersey 08544, USA.}%
 \affiliation{ 
Princeton Plasma Physics Laboratory, Princeton, New Jersey 08540, USA.}

\author{Kirill Lezhnin}%
 \affiliation{ 
Princeton Plasma Physics Laboratory, Princeton, New Jersey 08540, USA.}

\author{Michael J. Rosenberg}%
 \affiliation{ 
Laboratory for Laser Energetics, University of Rochester, Rochester, New York 14623, USA.}%

\author{S. X. Hu}%
 \affiliation{ 
Laboratory for Laser Energetics, University of Rochester, Rochester, New York 14623, USA.}%

\author{Dan Kalantar}%
\affiliation{ 
Lawrence Livermore National Laboratory, Livermore, California 94550, USA.}%

\author{Clement Trosseille}%
\affiliation{ 
Lawrence Livermore National Laboratory, Livermore, California 94550, USA.}%

\author{Hye-Sook Park}%
\affiliation{ 
Lawrence Livermore National Laboratory, Livermore, California 94550, USA.}%

\author{Bruce A. Remington}%
\affiliation{ 
Lawrence Livermore National Laboratory, Livermore, California 94550, USA.}%

\author{Gennady Fiksel}%
\affiliation{ 
Center for Ultrafast Optical Science, University of Michigan, Ann Arbor, Michigan 48109, USA.}%

\author{Dmitri Uzdensky}%
\affiliation{ 
Center for Integrated Plasma Studies, University of Colorado, Boulder Colorado 80309, USA.}%

\author{Amitava Bhattacharjee}%
 \affiliation{ 
Department of Astrophysical Sciences, Princeton University, Princeton, New Jersey 08544, USA.}%
 \affiliation{ 
Princeton Plasma Physics Laboratory, Princeton, New Jersey 08540, USA.}

\author{William Fox}%
 \affiliation{ 
Department of Astrophysical Sciences, Princeton University, Princeton, New Jersey 08544, USA.
}%
 \affiliation{ 
Princeton Plasma Physics Laboratory, Princeton, New Jersey 08540, USA.}

\date{\today}

\begin{abstract}
We present results from X-ray imaging of high-aspect-ratio magnetic reconnection experiments driven at the National Ignition Facility. Two parallel, self-magnetized, elongated laser-driven plumes are produced by tiling 40 laser beams. A magnetic reconnection layer is formed by the collision of the plumes. A gated X-ray framing pinhole camera with micro-channel plate (MCP) detector produces multiple images through various filters of the formation and evolution of both the plumes and current sheet. As the diagnostic integrates plasma self-emission along the line of sight, 2-dimensional electron temperature maps $\langle T_e \rangle_Y$ are constructed by taking the ratio of intensity of these images obtained with different filters. The plumes have a characteristic temperature $\langle T_e \rangle_Y = 240 \pm 20$ eV at 2 ns after the initial laser irradiation and exhibit a slow cooling up to 4 ns. The reconnection layer forms at 3 ns with a temperature $\langle T_e \rangle_Y = 280 \pm 50$ eV as the result of the collision of the plumes. The error bars of the plumes and current sheet temperatures separate at $4$ ns, showing the heating of the current sheet from colder inflows. Using a semi-analytical model, we find that the observed heating of the current sheet is consistent with being produced by electron-ion drag, rather than the conversion of magnetic to kinetic energy.
\end{abstract}

\maketitle

\section{\label{sec:introduction}Introduction}
Magnetic reconnection is a fundamental process whereby two plasmas carrying oppositely oriented magnetic field merge, driving the reconfiguration of the field-line topology inside a reconnection layer (also know as a current sheet)\cite{Biskamp2000,Zweibel2016}. As the magnetic lines break and reconnect, they induce out-of-plane electric field $\mathbf{E}$. Reconnecting systems are ubiquitous both in cosmic and laboratory plasmas. Thus, their dedicated investigation\cite{Yamada2016} has applications to our understanding of coronal mass ejections in the Sun\cite{Bhattacharjee1991,Edward1991}, solar flares\cite{Yan2022}, fast radio bursts\cite{Lyubarsky2020,Mahlmann2022}, gamma ray bursts\cite{Giannios2008,Mckinney2012}, blazars and active galactic nuclei jets\cite{Giannios2009}, sawtooth crashes in tokamaks\cite{Biskamp1994}, the radiative properties of black hole accretion disks \cite{Ripperda2020}, electron properties in the Earth's magnetosphere\cite{Burch2016,Phan2018}, pulsar magnetospheres\cite{Uzdensky2014,Cerutti2015,Philippov2018,Philippov2019}, and heat transport in hohlraums for indirect drive inertial confinement fusion\cite{Nilson2006}, amongst others. 

As magnetic reconnection occurs, the energy stored in the magnetic field is rapidly converted to thermal and bulk kinetic energy, heating up the plasma and accelerating it out of the current sheet\cite{Biskamp2000,Zweibel2016}. However, realistic plasmas (both in nature and the laboratory) can have a plethora of competing heating and cooling mechanisms that can potentially affect the reconnection process (and vice versa), such as collisional viscous heating, radiative cooling, and heat conduction. Therefore, it is important to probe the structure and evolution of the fluid during the reconnection process, ultimately to understand how the plasma and magnetic field co-evolve as electromagnetic energy is converted to fluid internal and bulk energy, and in turn, how it interplays with other plasma processes. Laboratory experiments allow investigating the energy equipartition in controlled reconnecting systems.

In this paper, we present results from magnetic reconnection experiments conducted at the National Ignition Facility (NIF). The goal is to measure the electron temperature in the laser-driven plumes and reconnection layer simultaneously. We aim at determining if there is significant electron heating in the layer and, if so, quantify the relative strength of heating and cooling mechanisms that may play a role in the observed temperature profiles. A quasi-2D, highly-extended reconnection geometry is produced by driving the reconnecting plasmas using multiple drive laser beams. As shown in Figure \ref{fig:exp_setup}, the spatially and temporally resolved plasma self-emission is observed using a gated X-ray camera with micro-channel plates with the line of sight aligned with the reconnecting magnetic field. The plasma self-emission is filtered with thin foils, and imaged using pinholes. The ratio between the images obtained provides a measure of the electron temperature averaged along the line of sight\cite{Schaeffer2021}. We find that the electrons in the current sheet are heated from the interaction between the laser-driven plumes. We use a semi-analytic model to quantify the relative importance of different heating/cooling mechanisms, including magnetic reconnection, electron-ion collisional drag, adiabatic expansion, and radiative cooling. Our calculations indicate that the observed heating does not result from flow stagnation (i.e., conversion from ion bulk to thermal energy), nor does magnetic reconnection significantly contribute. Instead, electron heating is likely due to electron-ion collisional drag in the interaction region.

Our analysis is based on statistically assembling temperature maps calculated independently from redundant filtered self-emission images, a technique which allows identification of systematic and random errors. We show that plasma expansion at early times ($t \approx 2$ ns) can lead to inconsistent temperature measurements if these errors are not accounted for. We also give an estimate on how to deal with systematic errors to give consistent results. However, at later times, the measurement uncertainty is mainly due to finite photon statistics which produce a speckled temperature pattern on the inferred maps. Therefore, the total uncertainty propagation can be assessed through an overall statistical ensemble which averages over the random error. Thus, combining multiple measurements on the same shot enables a statistical analysis to assess the measurement uncertainty without relying solely on a photometric model. 

This article is structured as follows: Section \ref{sec:methods} presents the experimental setup and analysis methods, and Section \ref{sec:results} shows the raw experimental images and post-processed temperature maps. We discuss potential electron heating mechanisms of the reconnection layer in Section \ref{sec:discussion}, and present conclusions and scope for future work in Section \ref{sec:conclusions}.


\section{\label{sec:methods}Experimental setup and methods}

\subsection{Experimental setup}
\begin{figure}
    \centering
    \includegraphics[width=8.5cm]{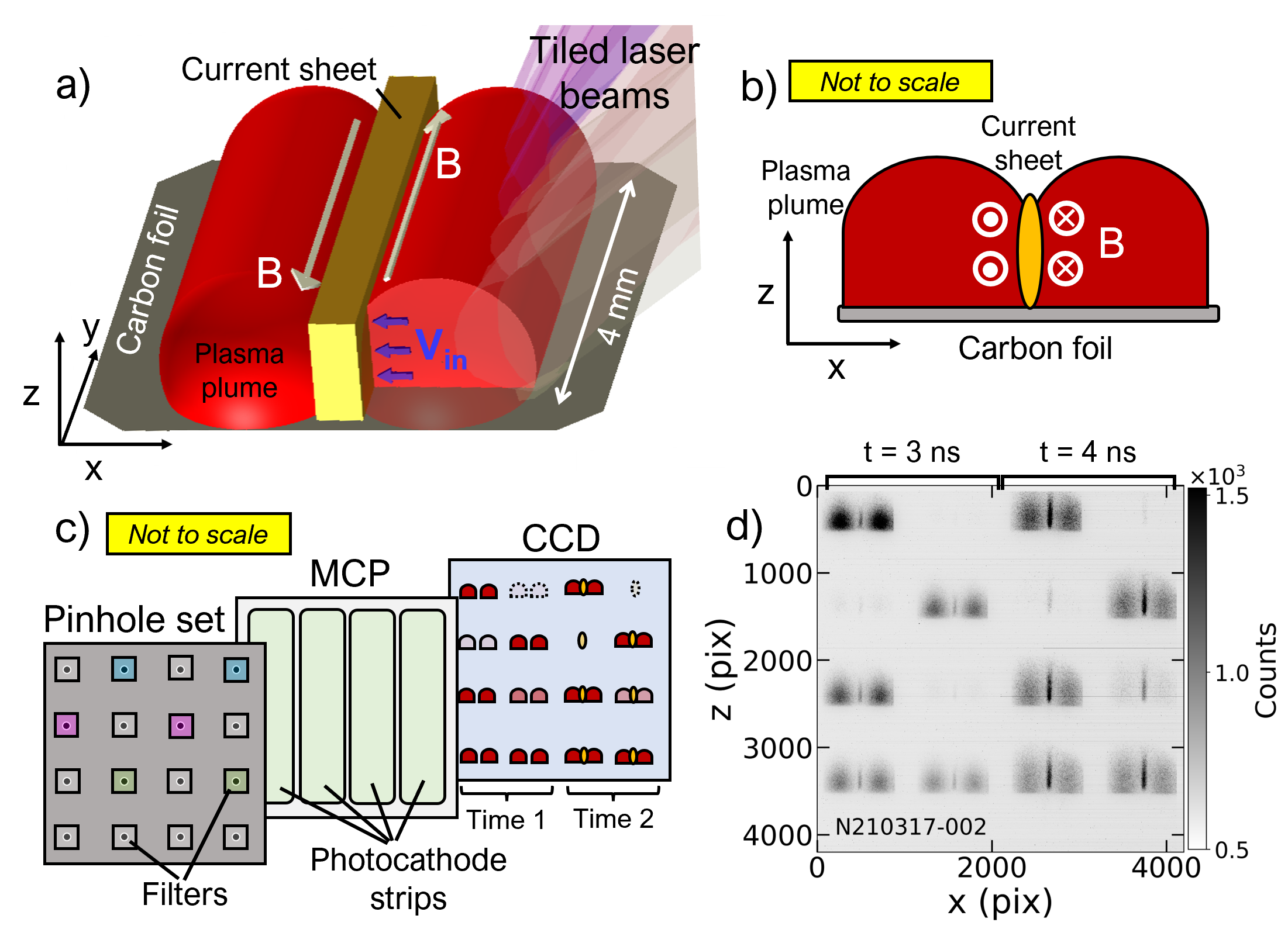}
    \caption{Experimental setup. a) Schematic 3-D illustration of high-aspect-ratio magnetic reconnection experiment. Only one side of the tiled beams is shown. b) Schematic of plasma dynamics from the GXD line of sight. c) Schematic of key GXD components (pinholes, filters, MCP, and CCD), and timing configuration. Different coloured squares on the pinhole set denote different filters fielded in the experiment. d) Example of raw X-ray experimental images.}
    \label{fig:exp_setup}
\end{figure}

In the experiments, a high-density carbon foil (5.5 $\times$ 6 mm$^2$, 15 $\mu$m thickness) was driven by a total of 40 beams (351 nm wavelength, 600 ps square pulse, 100 J/beam, $10^{14}$ Wcm$^{-2}$ total intensity) from the bottom outer cones of the NIF\cite{Fox2020}. Tiling sets of twenty beams produces a pair of 1-mm width by 4-mm length laser spots with a lateral separation of 2.4 mm, generating two elongated parallel plasma plumes, as shown in Figure \ref{fig:exp_setup}a. The plasma plumes expand and become magnetized by the Biermann battery effect\cite{Yates1982,Fiksel2014,Matteucci2018}. As these plumes expand inwards with velocity $V_{\text{in}}$, they collide in the mid plane and drive magnetic reconnection in the current sheet\cite{Fox2020}, as illustrated in Figure \ref{fig:exp_setup}a. The extended current sheet configuration makes the experiment quasi-2-dimensional. We use the NIF gated X-ray detector to obtain images on the out-of-target line of sight, spatially resolving the plumes and current sheet, as shown in Figure \ref{fig:exp_setup}b. Henceforth, we will neglect the effects of curvature and choose coordinates such that the inflow velocity $\mathbf{V_{\text{in}}}$ is along $\mathbf{\hat{x}}$ and the magnetic field $\mathbf{B}$ along $\mathbf{\hat{y}}$.

\begin{table*}
\caption{\label{tab:filters}Summary of filter sets and DC biases on the GXD camera (RGXD4F) setup for each experiment.}
\begin{ruledtabular}
\begin{tabular}{lcccccc}
 Shot ID&Gate timing (ns)&DC bias (V)&Filter 1&Filter 2&Filter 3&Filter 4\\ \hline
 N191114-001\footnote{Only one filter type was applied for each MCP strip in this experiment, as described by Schaeffer et al.\cite{Schaeffer2021}.}
 &2&+300&3 $\mu$m Al&6 $\mu$m Al&--&--\\
 N210317-002&3&+300&3 $\mu$m Al&$4.5$ $\mu$m Al&3 $\mu$m Al$+12.5$ $\mu$m Be&3 $\mu$m Al$+25$ $\mu$m Be\\
  N210317-002&4&+250&3 $\mu$m Al&$4.5$ $\mu$m Al&3 $\mu$m Al$+12.5$ $\mu$m Be&3 $\mu$m Al$+25$ $\mu$m Be\\
 N201117-004&5&+200&3 $\mu$m Al&$4.5$ $\mu$m Al&3 $\mu$m Al$+12.5$ $\mu$m Be&3 $\mu$m Al$+25$ $\mu$m Be\\
  N201117-004&6&+150&3 $\mu$m Al&$4.5$ $\mu$m Al&3 $\mu$m Al$+12.5$ $\mu$m Be&3 $\mu$m Al$+25$ $\mu$m Be\\
 N210317-001&7&+100&3 $\mu$m Al&$4.5$ $\mu$m Al&3 $\mu$m Al$+12.5$ $\mu$m Be&3 $\mu$m Al$+25$ $\mu$m Be\\
  N210317-001&8&+50&3 $\mu$m Al&$4.5$ $\mu$m Al&3 $\mu$m Al$+12.5$ $\mu$m Be&3 $\mu$m Al$+25$ $\mu$m Be\\
\end{tabular}
\end{ruledtabular}
\end{table*}

\subsection{GXD configuration}

A gated X-ray detector\cite{Oertel2006} (GXD) with a microchannel plate (MCP) was used to record images of the X-ray plasma self-emission in the $0.2 - 10$ keV spectral range. The GXD camera was located on the diagnostic instrument manipulator (DIM) 90-78 and its line of sight aligned with the plumes, as shown in Figure \ref{fig:exp_setup}b. 

The GXD camera has four rectangular (7.5$\times$36 mm$^2$)  MCP photocathode strips which can be triggered independently, and a phosphor screen coupled to a CCD, as shown in Figure \ref{fig:exp_setup}c. The gain of each strip was controlled by applying a DC bias voltage. The MCP strips are gated using a sweeping 600 ps mode, $\sim 3$ kV amplitude pulse with a $\sim 250$ ps transit time across the strip. Four images of the experiments were formed on each strip using sixteen-pinhole sets located at 478 mm from target chamber center and 836 mm from the MCP, yielding a magnification $M = 1.75$. The sweeping gate propagates at $\sim 140$ mm/ns, so a 2 mm image will have a skew across it of $\sim 15$ ps. This skew is much smaller than the $230$ ps integration time from the 600 ps electrical pulse. Therefore, the diagnostic time resolution is short compared to the plasma dynamical timescales ($\gtrsim 1$ ns). Indeed, as shown in Schaeffer et al.\cite{Schaeffer2021}, the electron temperature (within experimental uncertainty) is constant within the transit time of 250 ps across the strip. A single plasma image is recorded with an integration time $\lesssim 230$ ps, setting the imaging time resolution, which is also shorter than the timescales over which the plasma appreciably evolves. 

Therefore, all images from a single strip effectively probe the same snapshot of the plasma evolution. Each pinhole was $a = 150$ $\mu$m diameter, therefore the spatial resolution of the diagnostic is $a(1+M)/M\sim 240$ $\mu$m at the object plane. The MCP converts the X-ray photon signal into electrons, which are subsequently converted to visible light by a phosphor screen, and the resulting images are recorded on an SI-1000 CCD camera.

The plasma X-ray self-emission on each image was filtered using thin filter foils placed behind each pinhole. The diagnostic was set up such that each strip contains four images with three different filters ranging from a soft filter (3 $\mu$m Al) to a hard filter (25 $\mu$m Be $+$ 3 $\mu$m Al). Building on previous work by Schaeffer et al.\cite{Schaeffer2021}, the filter configuration is such that pairs of strips have the same bias and gate, but a diverse filter selection. We set up the timing such that two adjacent strips were co-timed, as shown in Figure \ref{fig:exp_setup}d. Moreover, this filter configuration allows obtaining redundant line-averaged temperature measurements which can be used to constrain the experimental uncertainty. Table \ref{tab:filters} summarizes the filter selection together with the applied DC voltage fielded in each experiment. 

\subsection{Analysis method \label{analysisSection}}

At electron temperatures $T_e \gtrsim 100$ eV and a broad range of densities, carbon atoms become fully ionized\cite{Chung2005}. Therefore, the plasma spectral emissivity $j_\nu$ (power density per unit frequency) is dominated by continuum bremsstrahlung emission and given by\cite{Hutchinson2002}
\begin{equation}\label{eq:emissivity}
    j_\nu = \frac{e^6}{(4\pi)^4\epsilon_0^3} \frac{32 \pi^2}{3\sqrt{3}m_e^2 c^3} \left( \frac{2m_e}{\pi} \right)^{1/2} \bar{g} \frac{Z^3 n_e^2}{\sqrt{T_e}} \exp{\left[\frac{-h\nu}{k_BT_e}\right]},
\end{equation}
where $e$ is the fundamental charge, $\epsilon_0$ is the permittivity of vacuum, $m_e$ is the electron mass, $c$ is the speed of light, $Z$ is the plasma charge state, $n_e$ is the electron number density, $\bar{g}$ is the Gaunt factor, and $h$ and $k_B$ are Planck's and Boltzmann's constants, respectively. Given the plasma composition, its density and temperature, we expect it to be optically thin \cite{Rybicki2004}. Therefore, the imaging system retrieves information about the weighted average plasma emissivity intensity integrated along the line of sight 
\begin{equation}
    \langle j_\nu \rangle_Y(x,z) = \int_{L_P} j_\nu(x,y,z) dy /L_P,
\end{equation}
where $L_P$ is the characteristic plasma length along the line of sight. If the MCP detector has a spectral response function $K=K(\nu)$, the ratio of intensity between two images (call them $I_1$ and $I_2$) can be obtained with filters with spectral transmission functions $W_1 = W_1(\nu)$ and $W_2 = W_2(\nu)$. Thus, the intensity ratio between corresponding pixels in each image is given by\cite{Schaeffer2021}
\begin{equation}\label{eq:ratio}
    \frac{I_1}{I_2}(x,z) = \frac{\int_0^{\infty} \langle j_\nu \rangle_Y K(\nu) W_1(\nu) d\nu}{\int_0^{\infty} \langle j_\nu \rangle_Y K(\nu) W_2(\nu) d\nu},
\end{equation}
and is determined only by the electron temperature because $\nu$ is coupled only to the $T_e$ in the exponential term of equation (\ref{eq:emissivity}). Since the GXD is located at 1.3 m from the plasma and the images have sizes of a few centimeters across, point-projection effects are of order $\sim 1\%$ and can be ignored. Hence, equation (\ref{eq:ratio}) can be inverted numerically at each pixel to solve for the electron temperature spatially averaged along the line of sight $\langle T_e \rangle_Y$. Based on pinhole diameter and filter thickness tolerance, we estimate an accuracy of $\sim 10$ eV (random error), which sets the minimum measurement uncertainty. The image processing is explained in detail in Appendix \ref{app:HIPPIE}.


\section{\label{sec:results}Results}
\subsection{Plasma formation and dynamics}

\begin{figure*}
    \centering
    \includegraphics[width=18cm]{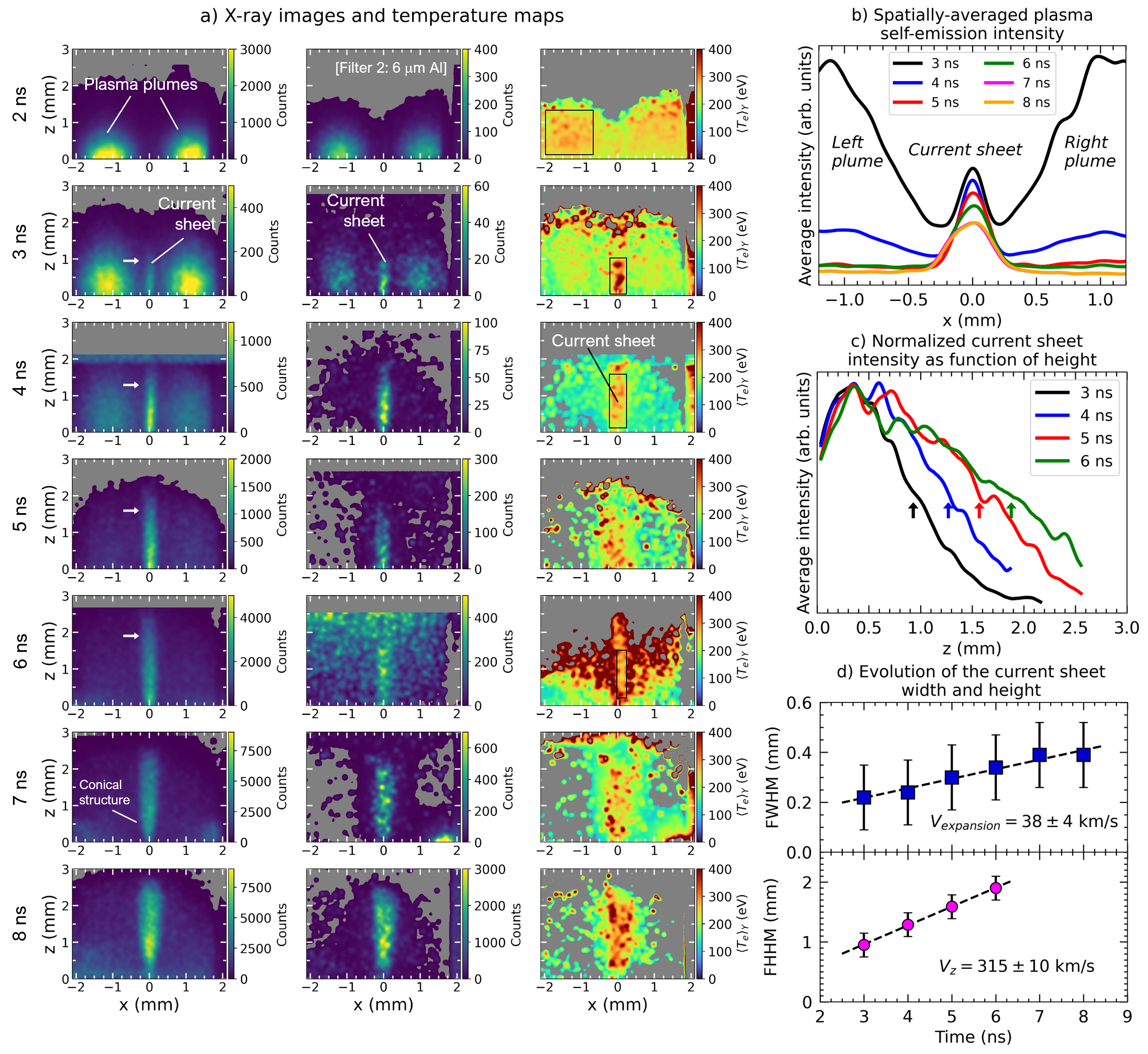}
    \caption{X-ray imaging characterization of the plasma dynamics. a) Self-emission images at different timings for filters 1 and 3 (c.f. Table \ref{tab:filters}) and corresponding line averaged electron temperature map; with the exception of data at 2 ns, which used filters 1 and 2. The coordinate system is defined with $z=0$ corresponding to the upper edge of the target. Masked grey areas correspond to regions at signal floor level on the CCD. Stray light is visible for Filter 3 at 6 ns. To help with visualization, we have masked that area in the calculated temperature map. Overall changes in the absolute intensity at different times are due to changes in the applied DC bias voltage, rather than an inherent irreproducibility of the plasma experiment. b) Horizontal lineouts of the spatially averaged intensity from Filter 1 images. Intensity was re-scaled to aid with the visualization. c) Spatially averaged and normalized lineouts of the current sheet self-emission. Coloured arrows indicate the upper end of the current sheet, also indicated in white arrows in panel (a). At times $t \geq 7$ ns, the current sheet length decreases likely due to reduced intensity associated with an overall decreasing density and have been excluded from the analysis. d) Evolution of the current sheet's width and length, the latter indicated by the arrows in panels (a) and (c). Top panel: the error bars of the FWHM are given by the $240$ $\mu$m spatial resolution. Bottom panel: the error bars contain contributions of both spatial resolution and image blurring introduced by the time streak.}
    \label{fig:plasma_evolution}    
\end{figure*}

Figure \ref{fig:plasma_evolution} shows X-ray self-emission datasets which characterize the expansion of the plasma plumes, subsequent collision, and current sheet formation and evolution. Panel (a) shows a collection of self-emission images obtained from two different filters, together with the calculated electron temperature map (which is discussed in Section \ref{subsec:temperature} below). Due to the scaling $\langle j_\nu \rangle_Y\propto \int n_e^2(x,y,z) dy$, the intensity of each image contains qualitative information about the plasma density, which can be interrogated and used as a proxy to estimate the overall plasma structure and its evolution. 

At $t=2$ ns, the plasma plumes have a characteristic size of $1$ mm, comparable to the laser irradiation pattern, and consequently remain spatially separated. The current sheet is formed and visible in the X-rays in the mid plane once the plumes merge at the collision time $t = t_{\text{col}} \equiv 3$ ns. The driver beams have a $1.2$ mm spatial full width at half maximum (FWHM). Assuming that the first colliding fluid elements were created at the location of the laser beam FWHM, then their distance to the mid plane is $D = 700 \pm 240$ $\mu$m, and the inflow velocity can be estimated as $V_{\text{in}} \sim D/t_{\text{col}} = 230 \pm 65$ km/s, in agreement with velocity estimates obtained from proton imaging\cite{Fox2020}. 
 
The reconnection layer forms close to the target at $t \lesssim 3$ ns with an initial height $H \lesssim 1$ mm and thickness $2\delta = 0.23 \pm 0.13$ mm (it is common to describe the current sheet in terms of its half-thickness $\delta$). Since this initial width is comparable to the diagnostic spatial resolution, it is possible that the current sheet is smaller that the self-emission region and its image is blurred due to the finite pinhole size. Indeed, proton radiography\cite{Fox2020} has shown a reconnection layer as small as $2\delta \sim 30$ $\mu$m. The layer remains stable and collimated throughout the experimental time frame.

Figure \ref{fig:plasma_evolution}b shows intensity lineouts averaged over the vertical extent of the plasma. The results show how the plume self-emission intensity is symmetric about the axis and decreases in time relative to the current sheet. Panel (c) shows normalized vertical lineouts of the current sheet self-emission intensity for different times. The lineouts show steady broadening, indicating that the reconnection layer slowly widens whilst also expanding axially throughout the plasma evolution. 

To estimate the current sheet's transverse expansion velocity $V_{expansion}$, the characteristic current sheet width was defined as the FWHM of each lineout in panel (b). As shown in Figure \ref{fig:plasma_evolution}d, the data shows a linear trend associated with the slow expansion of the reconnection layer from which we calculate $V_{expansion} = 38 \pm 4$ km/s $\ll V_{\text{in}}$, which shows that the current sheet exhibits steady-state behaviour in the inflow direction. Similarly, the time series of the current sheet's height, defined as the full height at half maximum (FHHM), can be used to estimate the vertical velocity to be $V_z = 315 \pm 10$ km/s (see Figure \ref{fig:plasma_evolution}c and d). At these velocities we expect the blurring along the $Z-$axis due to the time resolution to be $\sim 100$ $\mu$m, which is comparable to the spatial resolution, and therefore has been considered in the error bar of the FHHM by adding it in quadrature to the $120$ $\mu$m half spatial resolution for a total $200$ $\mu$m uncertainty.

We also notice that at $t = 7$ ns, the reconnection layer starts losing collimation and develops into a conical-like structure. At $t=8$ ns, the divergence angle is $\sim 14^{\circ}$ and may be an indication that the plumes have been mass-depleted, and therefore cannot provide sufficient pressure to confine the current sheet, leading to the onset of plasma disassembly.

\subsection{Spatially-averaged temperature measurements}\label{subsec:temperature}

As discussed previously, the intensity ratio between images can be used to infer the electron temperature averaged along the line of sight. However, using equation (\ref{eq:ratio}) requires making a series of assumptions to treat the datasets, namely that the images are obtained under the same MCP gain on both strips, and that the pinhole diameters and filter thicknesses are both reproducible and accurate to experimental specification. In addition, it is important to assess the effect of finite photon statistics, which may lead to inhomogeneous temperature patterns that do not correspond to inherent properties of the plasma. In this work, we have accounted for random sources of uncertainty by statistically averaging over multiple image pairs, yielding temperature measurements with well-characterized uncertainties.

\begin{figure}
    \centering
    \includegraphics[width=8.5cm]{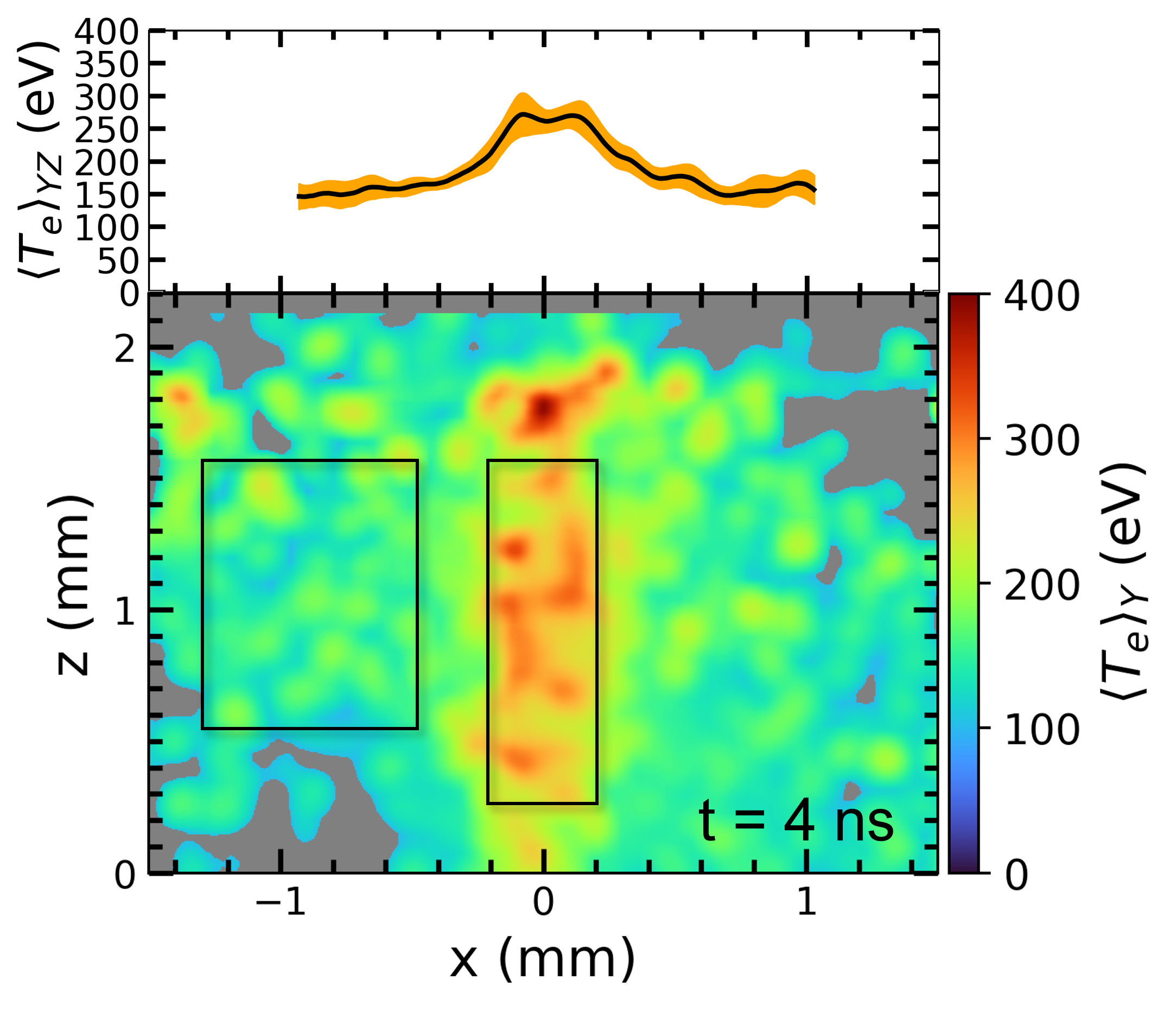}
    \caption{Line-averaged electron temperature map at $t=4$ ns. Boxes  represent domains where spatial averages are calculated in the regions of interest (plumes and current sheet). Upper panel: vertically averaged electron temperature ($\langle T_e \rangle_{YZ} \propto \int \langle T_e \rangle_Y dz$). Orange shade shows the $1\sigma$ spatial standard deviation.}
    \label{fig:single plasma image}
\end{figure}

 Figure \ref{fig:single plasma image} shows a typical temperature map of the interacting plasma. We defined "plume" and "current sheet" regions as indicated. The self-emission intensity emitted from the plumes is strong enough to image them together with the current sheet up to $t=4$ ns. After that, the self-emission intensity from the plumes becomes too low and falls outside the dynamic range of the diagnostic and only the current sheet is visible. The plumes exhibit a homogeneous temperature value, with no significant large scale temperature gradients. However, the higher filtered images presented in Figure \ref{fig:plasma_evolution}a show speckled features due to the lower photon counts. These patterns are correlated with similar ones in the inferred temperature maps with characteristic sizes below the imaging spatial resolution. Therefore, these are artifacts that originate from finite photon statistics rather than real properties in the plasma. These features complicate the analysis because they create patches with zero or divergent emission ratios, and therefore non-physical temperatures (via equation (\ref{eq:ratio})). To address this, we applied a Gaussian smoothing with FWHM $= 115$ $\mu$m in the object plane before aligning and taking the ratio between the intensity on pixels on the respective images. The smoothing function acts within the diagnostic $\sim 240$ $\mu$m spatial resolution and ensures consistency in the analysis while also retaining information about the characteristic variance in temperature measurements obtained from the photometry. We have checked that the smoothing does not significantly change the inferred characteristic value of $\langle T_e \rangle_Y$. The smoothing FWHM of 115 $\mu$m was chosen as the minimum value to obtain numerical results for the regions of interest across all images and that is simultaneously well below the spatial resolution. Stronger smoothing would result in the underestimation of the uncertainty of the characteristic temperature.

To obtain the characteristic temperature of the plasma (plumes and reconnection layer), we take representative samples from the region of interest of each temperature map and calculate the average and standard deviation of the temperature. Examples of these regions are indicated using black rectangles in Figure \ref{fig:plasma_evolution}a. In the next section, we discuss the images and temperature maps in more detail for two representative timings.

\subsubsection{Weighted statistical ensemble of $\langle T_e\rangle_Y$ maps}

The GXD configuration allows obtaining up to 8 images per timing per experiment, which can be processed to obtain multiple line-averaged temperature maps. To combine the information between them, we took spatial averages of the inferred temperature of the plasma's regions of interest, namely inflowing plumes and current sheet (Figure \ref{fig:single plasma image}). This can be used to obtain ensemble-averaged values of temperature and assess the main sources of experimental uncertainty as well.

For each individual image, we define the spatially averaged electron temperature as
\begin{equation}\label{eq:spatial av}
    \langle T_e \rangle \equiv  \langle T_e \rangle_{XYZ} = \frac{1}{N_x N_z} \sum_{j}^{N_z}\sum_i^{N_x} \langle T_e \rangle_Y (x_i,z_j) ,
\end{equation}
where, for notation compactness, we have dropped the subscript information about the coordinate over which we average, $N_x$ and $N_z$ are the number of pixels along $x$ and $z$, respectively, inside the large representative regions (see Figure \ref{fig:single plasma image}), and $(x_i, z_j)$ are pixel coordinates where $\langle T_e \rangle_Y$ is evaluated. We note that the line-average along $Y$ is intrinsically set by the diagnostic line of sight, however integration in $XZ$ corresponds to the analysis coarse-graining. We have made sure that we average over large enough domains, such that the spatially averaged electron temperature is domain-size independent within the spatial standard deviation given by
\begin{equation}\label{eq:spatial sigma}
    \delta T_e = \left[ \frac{ \sum_{j}^{N_z}\sum_i^{N_x} \left( \langle T_e \rangle_Y (x_i,z_j) - \langle T_e \rangle \right)^2 }{(N_x -1)(N_z -1)}\right]^{1/2}.
\end{equation}

Further analysis is done after calculating $\langle T_e \rangle$ and $\delta T_e$ for the regions of interest in each $\langle T_e \rangle_Y$ map. For each timing a set of maps can be constructed, from which a weighted statistical ensemble can be calculated. The ensemble contains information of all images within a given set. The spatially-averaged electron temperature for a given ensemble is calculated using
\begin{equation}\label{eq:weighted spatial av}
    \bar{T}_e = \frac{\sum_j \omega_j \langle T_e \rangle_j} {\sum_j \omega_j},
\end{equation}
where the weight for the $j-$th data point with spatially-averaged temperature $\langle T_e \rangle_j$ is given by $\omega_j = (\delta T_{e})_j^{-1}$. The uncertainty of a given set of measurements is given by the weighted standard deviation
\begin{equation}\label{eq:weighted spatial sigma}
    \Delta T_e = \left[ \frac{N}{N-1} \frac{ \sum_j \omega_j(\bar{T_e}-\langle T_e \rangle_j)^2}{\sum_j \omega_j} \right]^{1/2},
\end{equation}
where $N$ is the number of images in the ensemble. Henceforth, this approach results in two distinct kinds of error bars. For each line-averaged temperature map, the spatially-averaged values are given by $\langle T_e \rangle$ with $1\sigma$ uncertainty given by $\delta T_e$. For each timing, a set of spatially average values calculated from a set of temperature maps has an ensemble average given by $\bar{T}_e$ and $1\sigma$ weighted uncertainty of the whole ensemble given by $\Delta T_e$. 

Gated MCPs are known to exhibit variable sensitivity along the photo-cathode strips characterized by a "droop" function\cite{Holder2016}. In addition, although operated at the same bias conditions, different strips can exhibit discrepancies in the overall gain. To properly assemble the images, we need to check that there are negligible systematic errors from the place on the detector where images are produced. We show below (Sections \ref{subsec:2ns} and \ref{subsec:4ns}) that finite photon statistics effects dominate over these sources of uncertainty.

\subsubsection{Plume temperature at 2 ns}\label{subsec:2ns}

\begin{figure*}
    \centering
    \includegraphics[width=17cm]{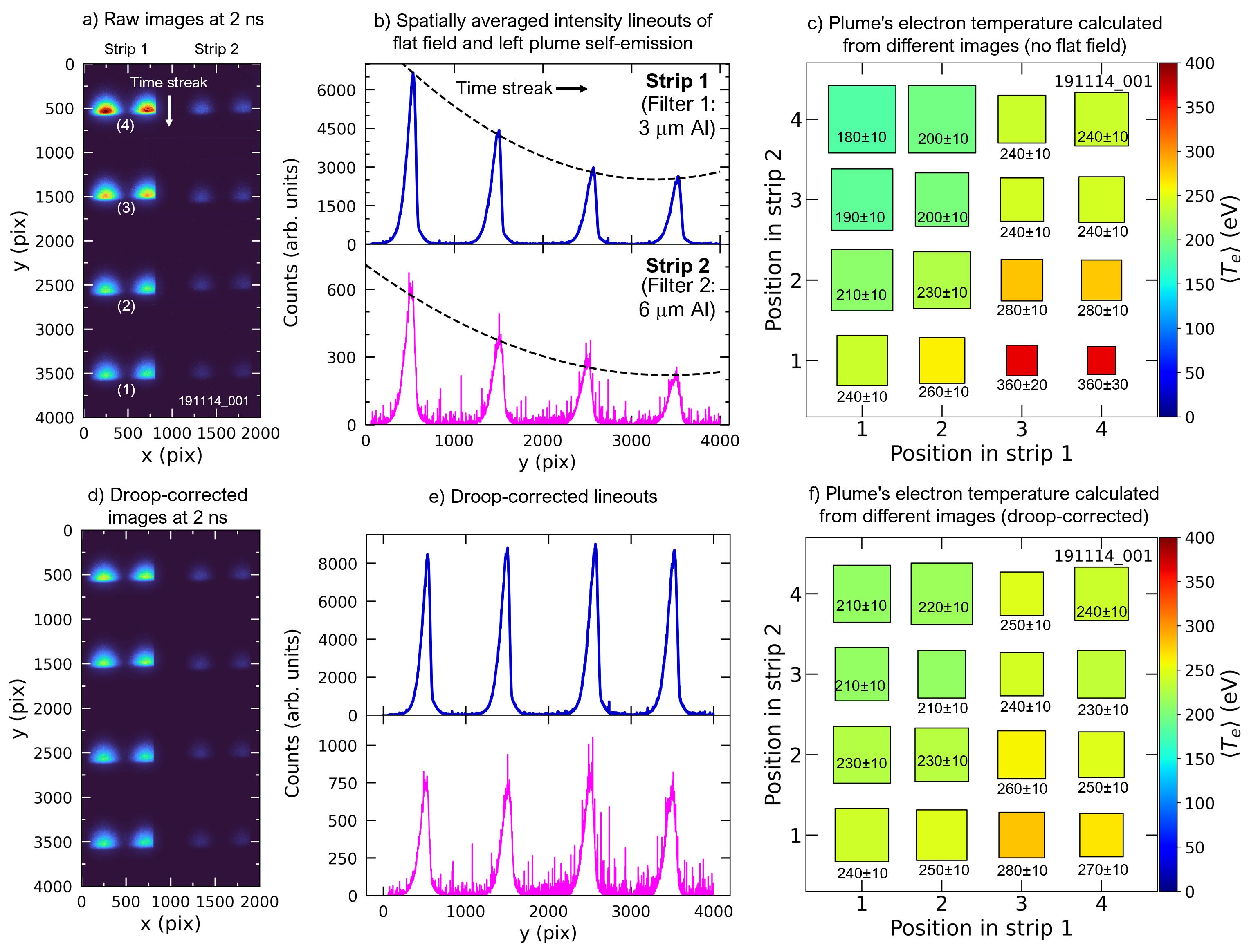}
    \caption{X-ray self-emission images and analysis of plasma plumes at $t=2$ ns. a) CCD image of the MCP strips 1 and 2. The sweeping gate moves down through the image. b) Intensity lineouts along strip 1 (upper panel) and strip 2 (bottom panel) averaged spatially in the $X$-axis around the left plume. Each strip was uniformly filtered with either $3$ $\mu$m Al foil (filter 1) or $6$ $\mu$m Al foil (filter 2). Dashed lines show parabolic fit of the local maxima which were used to infer the MCP droop function. c) Filter-filter diagram of the temperature calculation for the plumes using different pairs of images along each strip with no flat field (droop correction). The position is labelled 1 to 4 in the direction opposite to the time streak (also shown in panel (a)). d) CCD image corrected by droop function inferred from data. e) Same lineouts as panel (b) but after droop correction. f) Filter-filter diagram of the temperature calculation after droop correction.}
    \label{fig:single plume measurement}
\end{figure*}

In this section, we discuss the GXD configuration where images from a single filter type were located on each photo-cathode strip (as shown in Table \ref{tab:filters}). This is the case for the $t=2$ ns dataset only. Figure \ref{fig:single plume measurement}a shows a raw CCD image of the plasma plumes at $t=2$ ns. The two striplines contained in the image are labelled 1 and 2. The sweeping gate streaks downwards through the detector as indicated in the arrow.  Left and right hand side plumes are symmetric and independent analysis yields similar results (with discrepancies in their spatially averaged temperature $\lesssim 5$ eV). Hence, for simplicity we focus our analysis on the left plume. 

\begin{figure}
    \centering
    \includegraphics[width=8.5cm]{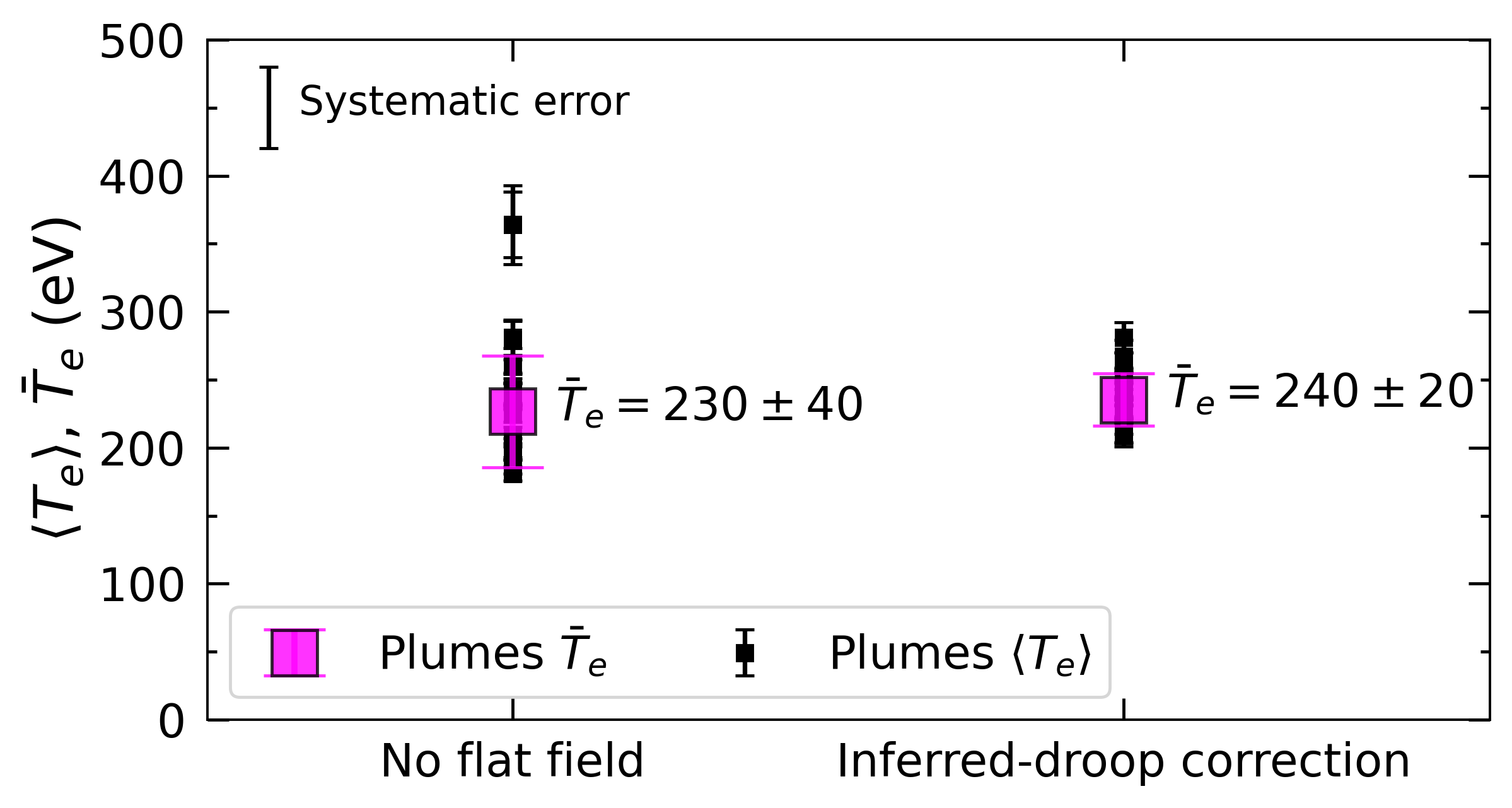}
    \caption{Comparison of inferred plume spatially-averaged temperature and statistical assembly for with and without flat field (droop) correction at $t=2$ ns. Black square markers indicate $\langle T_e \rangle$ on each image with $\delta T_e$ as error bar, whereas the magenta square $\bar{T}_e$ indicates the statistical average with $\Delta T_e$ as error bar (which are annotated).}
    \label{fig:single plume analysis comparison}
\end{figure}

The raw CCD images show the decreasing sensitivity of the detector along the direction of propagation of the gate, as the plasma intensity droops downwards. Lineouts of horizontally averaged intensity along each MCP stripline are presented in Figure \ref{fig:single plume measurement}b. We observe a systematic decrease in signal intensity of up to $60\%$. The top-to-bottom time of flight of the sweeping gate is $\sim 250$ ps, much shorter than any plasma time evolution which are noticeable on scales of nanoseconds. Therefore, we attribute the change in sensitivity to the MCP droop.

\begin{figure*}
    \centering
    \includegraphics[width=17cm]{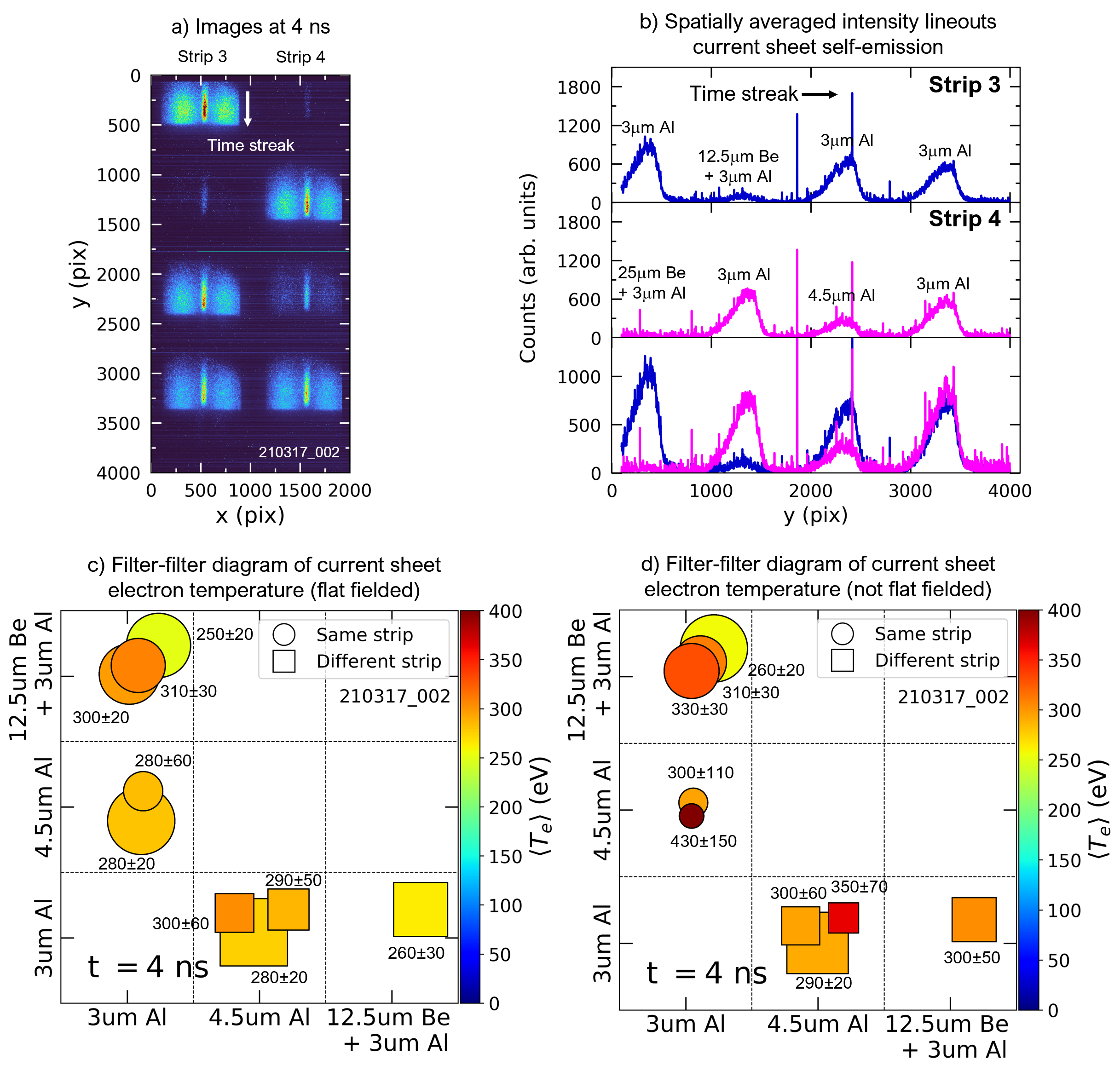}
    \caption{X-ray plasma self-emission images and analysis of current sheet at $t=4$ ns. a) CCD image of the MCP strips 1 and 2. The sweeping gate moves down through the image. b) Intensity lineouts of strip 3 (upper panel), strip 4 (middle panel), and both strips after flat field correction (bottom panel) around the current sheet. c) Filter-filter diagram of the temperature calculation for the current sheet using different pairs of images after flat field correction. Temperature values resulting from images on different strips are shown in squares at the bottom of the diagram, whereas images on the same strip are presented in circles on the top of the diagram. d) Same as panel (c) but without flat field correction.}
    \label{fig:current sheet measurement}
\end{figure*}

We investigated the effect of the self-emission variability along the detector on the inferred spatially-averaged electron temperature by analyzing all intensity ratios between every image pair. This process yields 16 different temperature maps. We have labelled each X-ray image on the CCD along the strip from the bottom up (i.e. opposite to the time streak), as convention. Figure \ref{fig:single plume measurement}c shows a diagram that summarizes the analysis result. Each marker color shows the spatially-averaged electron temperature $\langle T_e \rangle$. Their position in the horizontal and vertical axis indicate the images that were used to calculate the temperature map. Finally, the size of the marker is proportional to the spatial statistical weight $\omega$ (inverse of the spatial standard deviation).  In cases when the spatial variation $\leq 10$ eV, we have used the measurement accuracy\cite{Schaeffer2021} instead, i.e. $\delta T_e \geq 10$ eV. 

The diagonal components show a consistent $\langle T_e \rangle \sim 240$ eV. These correspond to side-to-side images on different strips and shows that the droop is similar along them. However, the inferred temperature gets systematically higher (and lower) when comparing the intensity of non-neighbouring images. The change is relatively symmetric about the diagonal, with a marginal additional contribution from high-fluence pixels in the temperature maps with higher average.

We can estimate the overall effect of the droop by inferring \textit{in situ} the droop function from the data and correcting accordingly. We fit parabolic functions to the peaks of plasma intensity lineouts, shown in dashed lines on Figure \ref{fig:single plume measurement}b, and use them to create the MCP droop function. We created a synthetic normalized flat-field image for both strips and corrected the plasma CCD image by dividing by the normalized flat-field. The result is shown in Figure \ref{fig:single plume measurement}d. The improvement is evident in the improvement of intensity consistency on the images. Lineouts at the same locations as panel (b) are shown in panel (c) which show the improvement in intensity consistency along the strip. We notice that the level of noise (photon statistics fluctuations) is exacerbated after the droop correction. This is because the fluctuation level is approximately constant in the raw images, and dividing by the monotonically decaying droop function amplifies it.

The equivalent diagram to panel (c) is shown in panel (f). We see a much improved measurement consistency when taking the ratio over all possible image pairs. There is still an inferred temperature gradient along the horizontal and vertical axis due to the increased inferred-temperature fluctuations.

Figure \ref{fig:single plume analysis comparison} compares the spatially-averaged temperature and ensemble average temperature resulting from both analyses (no flat fielding and flat fielding from inferred-droop). We see that the average changes marginally within the uncertainty given by $\Delta T_e$. Therefore, our temperature inference appears agnostic to the flat fielding process, as the result is resilient to the droop function (or lack thereof). 


In addition to the droop function, a secondary source of systematic error is given by discrepancies in gain between photo-cathode strips. By comparing images with equal filter and equal location on different strips in the 'checkerboard' pattern (from data corresponding to $t \geq 3$ ns), we found that even when set in equal configuration, the striplines can exhibit differences of up to $40\%$ in gain. When a single filter type is used on each strip, the effect of this discrepancy cannot be accounted for directly from the data. As shown in Appendix \ref{app:HIPPIE} (Figure \ref{fig:intensity ratio temperature}d), a change of $40\%$ in our ratio-range leads to 30 eV changes in the inferred temperature. However, we note that this is approximately the same as the error from spatial intensity modulations and therefore our conclusions are not sensitive to this within random uncertainty (see Figure \ref{fig:single plume analysis comparison}).

\subsubsection{Current sheet temperature at 4 ns} \label{subsec:4ns}

The data sets for timings $t > 2$ ns used a filter configuration with a `checkerboard' pattern such that strips 1 and 3 contain two filter types, and strips 2 and 4, three (see Figure \ref{fig:current sheet measurement}). Consequently, redundant intensity ratios can be taken with the same filters but with the self-emission being imaged on different parts of the GXD. Hence, images obtained on either the same or different striplines can be compared, which allows assessing the effect of uneven trigger levels on pairs on the photocathodes. In addition, the filter pattern allows more types of filters to be compared than with a single filter per strip configuration. A representative set of images is presented in Figure \ref{fig:current sheet measurement}a which contains data at $t = 4$ ns. Strip 3 has three images filtered with filter 1 and filter 2, whereas strip 4 has two images with filters 1, 2, and 4. We note that the images with the thickest beryllium foil are too dim to produce data, and therefore have been excluded for the rest of the analysis.

Figure \ref{fig:current sheet measurement}b shows horizontally-averaged intensity lineouts along the current sheet. Each bump on the lineout corresponds to a filtered image, with the corresponding filter annotated above. Similar to the previous section, we have used a droop function that decreases with the $Y$-axis to correct the CCD image. The droop-corrected lineouts are shown in Figure \ref{fig:current sheet measurement}b bottom panel. In addition, the gain difference between strips can be estimated from the intensity of images on the same position along the strip with the same filter, which is the case for the bottom images on Figure \ref{fig:current sheet measurement}a.

\begin{figure}
    \centering
    \includegraphics[width=8.5cm]{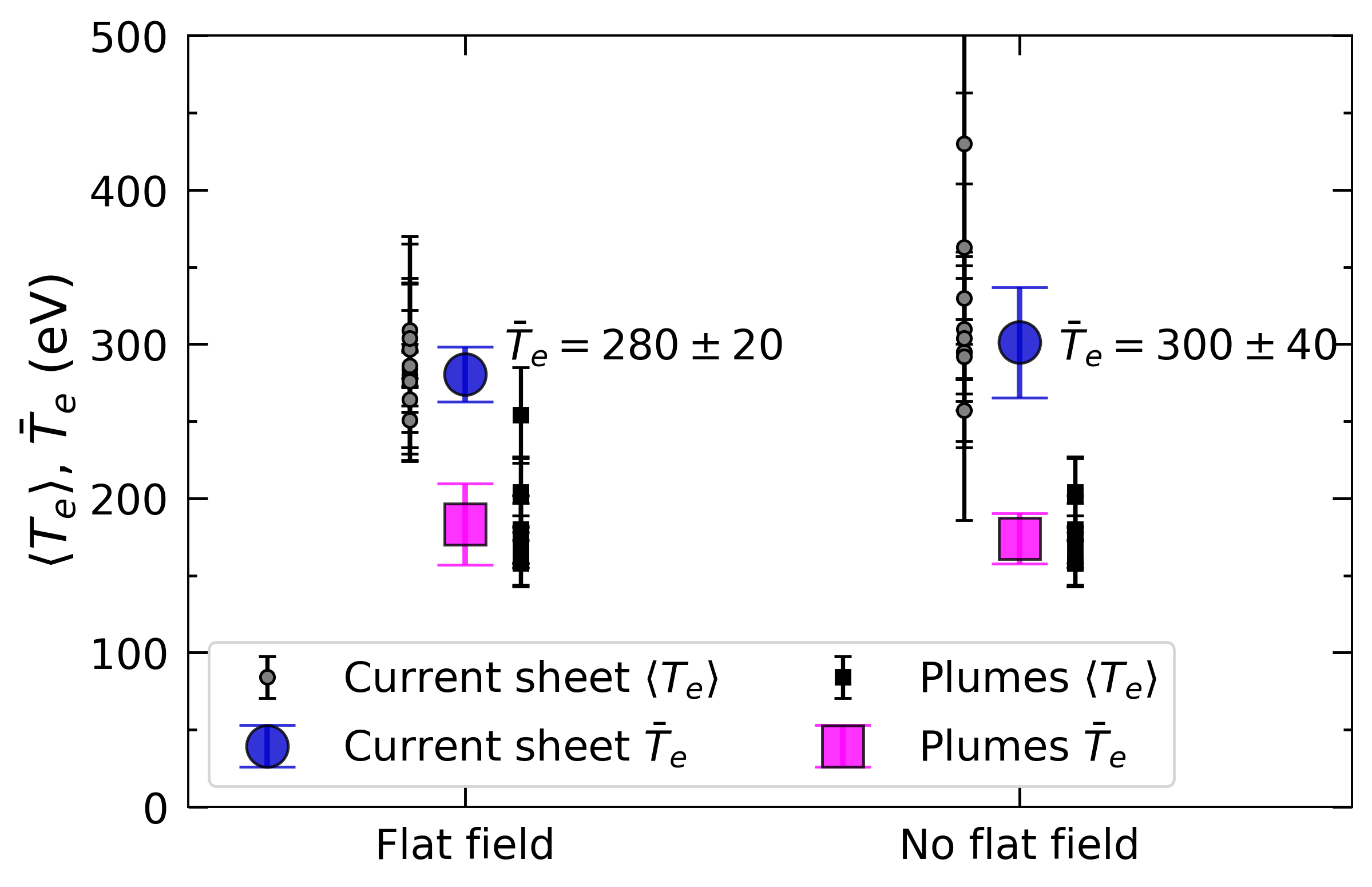}
    \caption{Comparison of inferred plasma spatially-averaged temperature and statistical assembly with and without flat field correction at $t=4$ ns. Black and grey markers indicate $\langle T_e \rangle$ on each image with $\delta T_e$ as error bar for the plumes and current sheet, respectively. Magenta and blue markers $\bar{T}_e$ indicate the statistical average with $\Delta T_e$ as error bar for the plumes and current sheet, respectively.}
    \label{fig:current sheet analysis comparison}
\end{figure}

We assess the effect of using different portions along MCP striplines, filter pairs, and different photo-cathode strips through filter-filter diagrams. Figure \ref{fig:current sheet measurement}c shows the result for the dataset after droop correction. Similarly to the diagrams in the section above, the markers' color indicates $\langle T_e \rangle$ and marker size is $\propto \delta T_e$. The horizontal and vertical axes show the filters that were used to obtain the spatially-averaged temperature. We note there are no diagonal terms because that would imply comparing intensity images through the same filters. We have differentiated between the measurements made by comparing images on the same strip (upper region, circles) vs. different strips (lower region, squares). Moreover, to avoid overlap, we have displaced the marker positions slightly. Systematic errors in the measurement from comparing images on strips with different gains or with a dynamically significant evolution along the sweeping gate would be reflected in drastic, localized inconsistencies in the diagram.

We find consistent spatially-averaged temperatures in the current sheet in the flat-field corrected images, with no strong systematics in the results. Figure \ref{fig:current sheet measurement}d is equivalent to panel (c) but removing the flat-fielding step in the analysis. We observe slightly less consistent values of $\langle T_e \rangle$, with a few outlying datapoints. Figure \ref{fig:current sheet analysis comparison} compares the results from these two analyses. We see that both for the plumes and current sheet, the ensemble average $\bar{T}_e$ does not change within the spatial standard deviation $\Delta T_e$, which shows that our results are independent of the flat-fielding step. This analysis was conducted for all datasets that used the 'checkerboard' pattern ($t=3$ to $8$ ns) and found no systematic discrepancies between intensity ratios from different filters and location on the GXD.

\subsubsection{Spatially-averaged electron heating and temperature evolution}

To summarize the analysis presented above:
\begin{itemize}
    \item[1.] For the $t=2$ ns data, a single-filter per GXD stripline was used. The results indicate that the droop function can lead to discrepancies in average temperatures; however, after ensemble-averaging, the inferred temperature is insensitive to the droop function. Nevertheless, the results can exhibit a $\pm 30$ eV systematic error emerging from strip-to-strip gain differences that cannot be accounted for in the configuration used.
    \item[2.] For all other datasets ($t\geq 3$ ns), the filter 'checkerboard' pattern can be used to assess the strip-to-strip gain difference and allows producing redundant temperature measurements. We found that there are no systematic errors from using different portions on the MCP.
    \item[3.] The main source of uncertainty is due to temperature spatial modulations that emerge from speckled self-emission patterns on the higher-filtered X-ray self-emission images, which lead to spatial modulations in the line-averaged temperature. 
\end{itemize}

\begin{figure}
    \centering
    \includegraphics[width=8cm]{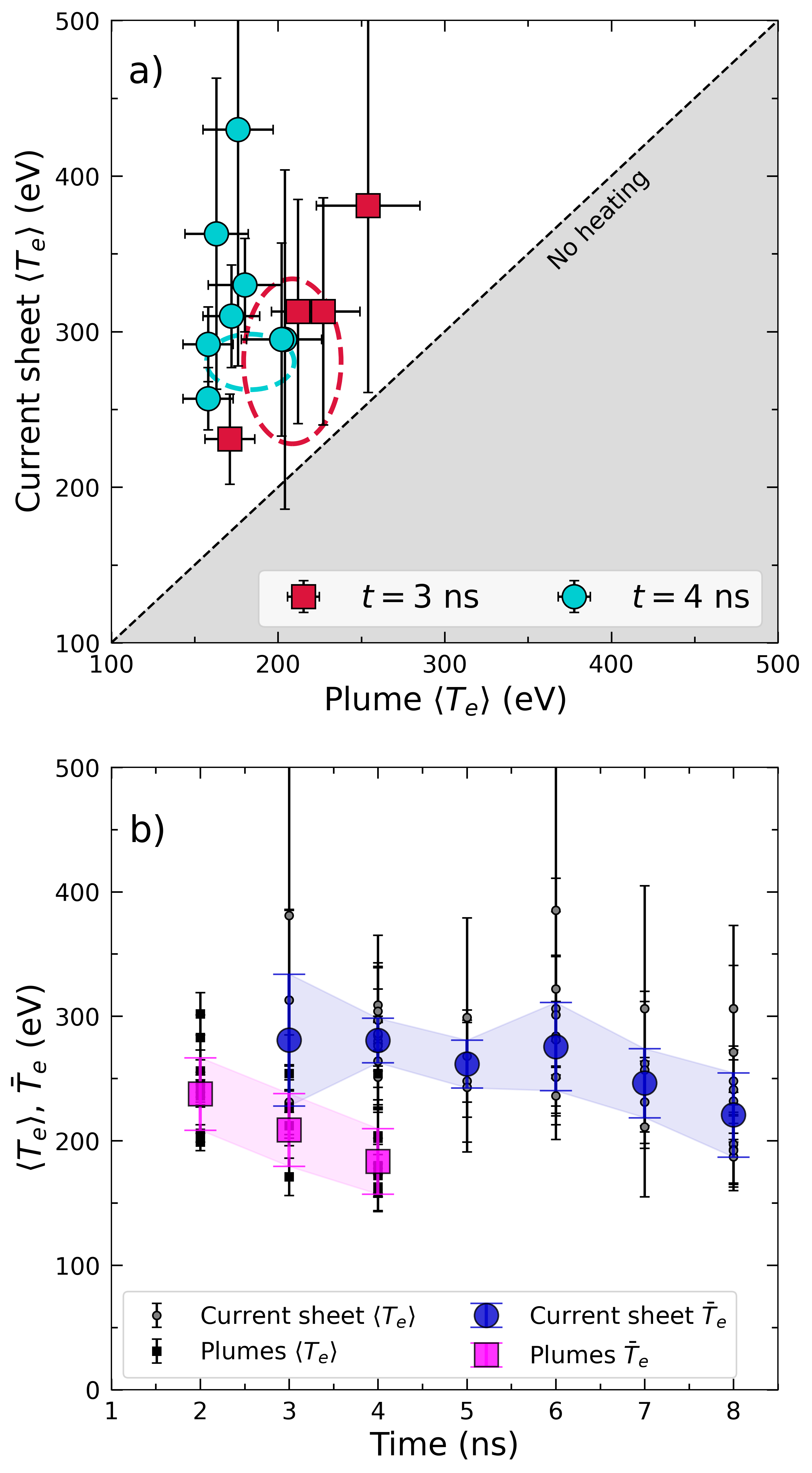}
    \caption{a) Scatter diagram of the current sheet's spatially-averaged electron temperature vs. the plume temperature for each $\langle T_e \rangle_Y$ image. The dashed ellipses are centred around the weighted average and the semi-minor and semi-major axes correspond to the weighted standard deviation of the plume and current sheet temperature, respectively. b) Spatially-averaged electron temperature time series. Error bars on black and grey data-points correspond to the spatial standard deviation $\delta T_e$, whereas error bars on blue and magenta data points correspond to the weighted standard deviation of the ensemble of images at each time. }
    \label{fig:temp_evolution}
\end{figure}

Electron heating in the current sheet can be assessed both from each individual image and through an ensemble average. Figure \ref{fig:temp_evolution}a shows the current sheet's spatially-averaged temperature versus the plumes' spatially-averaged temperature for the $t=3$ ns and $t=4$ ns datasets. The diagonal represents the no heating locus (i.e. thermal equilibration of the two regions). The horizontal and vertical error bars correspond to $\delta T_e$ in the plumes and current sheet, respectively. The data points are significantly above the diagonal, consistent with electron heating on each individual map. The dashed ellipses are centred around $\bar{T}_e$ with the semi-major and semi-minor axes corresponding to the $1\sigma$ uncertainty. As an ensemble, the $t=3$ ns shows marginal heating, however from the $t=4$ ns data, the electron heating in the current sheet is clear from colder inflowing plumes. 

To put the electron heating in the context of the overall evolution of the plasma, we have constructed a time series, shown in Figure \ref{fig:temp_evolution}b, from the droop-corrected measurements. Each point in grey corresponds to the average temperature $\langle T_e \rangle $ of the relevant feature (either plume or current sheet) and the error bar shows the $1\sigma$ spatial standard deviation $\delta T_e$ on that image, and therefore reflects statistical spatial fluctuations on the measurement. Colored markers (blue for the plumes and magenta for the current sheet) show the weighted average obtained from all possible image ratios.

The data suggests that the plumes exhibit a small amount of cooling for as long as they are detectable. In contrast, the current sheet shows a more slow decline. The plumes exhibit a temperature $\langle T_e \rangle = 240 \pm 20$ eV at $t = 2$ ns. The current sheet is formed at $t=3$ ns with a higher temperature $\langle T_e \rangle = 280 \pm 50$ eV, at which time the plumes' temperature drops slightly to $\langle T_e \rangle = 210 \pm 30$ eV, although they are within error bars. Therefore, the current sheet has a similar temperature as the plumes initially within measurement uncertainty. 

However, at $t=4$ ns, the uncertainty on the plumes and current sheet temperature is small enough such that it is clear that the reconnection layer has been heated compared to the inflowing plumes. The current sheet retains its temperature $\langle T_e \rangle = 280 \pm 30$ eV, as the plumes cool down slightly to $\langle T_e \rangle = 180 \pm 30$ eV. At later times, the plumes are too thin and/or cold and their luminosity is not high enough to be imaged. In contrast, the current sheet remains dense and hot, retaining its temperature for the duration of the experiment, except for a slight temperature decline after $t = 6$ ns, which corresponds to the development of the conical structure, loss of collimation, confinement, and therefore onset of the plasma disassembly.


\section{\label{sec:discussion}Discussion}

Our analysis detected significant heating of the current sheet, as we measured a $50\%$ electron temperature increase relative to the inflowing plasma plumes. Previous temperature measurements of laser-driven magnetic reconnection experiments have reported varied conclusions. For example, Nilson et al.\cite{Nilson2006} used the ion-acoustic wave (IAW) Thomson scattering channel at the Vulcan laser system and observed an increase of the electron temperature by a factor of $2$ in the current sheet a few hundred picoseconds after the driver beams were turned off. On the contrary, Rosenberg et al.\cite{Rosenberg2012} combined both IAW and electron plasma wave (EPW) channels of the Thomson scattering diagnostic at the OMEGA laser to measure temperature and found no significant heating of the reconnection layer. 

The difference between our results and theirs could be explained by fundamental differences between the plasma conditions under different drivers (i.e., OMEGA and NIF). As an example, experiments at the OMEGA Extended Performance (EP) laser with $0.5$ ns pulses found that energy was coupled to supra-thermal electrons (typical energies $> 100$ keV), rather than the bulk\cite{Fiksel2021}. However, it is also possible that using different diagnostics can lead to contradictory conclusions. Although a Thomson scattering diagnostic provides detailed measurements of plasma parameters such as density, temperature and bulk velocity, the scattered light is collected from a localized collection volume which can sample plasmas outside the current sheet. In fact, the collection volume usually has a characteristic diameter of $\sim 100$ $\mu$m, and therefore could be averaging over a plasma region larger than the reconnection layer. Layer widths of $30$ $\mu$m have been found in laser-driven magnetic reconnection experiments\cite{Fox2020}, which would contribute to missing the thermal decoupling. Moreover, given the small size of the current sheet, an accidentally mispointed probe beam could lead to missing the current sheet altogether in the absence of imaging diagnostics to use as cross calibration\cite{Suttle2021}.

Therefore, our results are in better agreement with previous work by Nilson et al.; though, the overall heating level in our experiments seems to be smaller than theirs. Since currently we do not have density measurements, the goal of this section is to constrain potential heating mechanisms that can account for the measured temperatures. This would help design future experiments as a follow up from this work. We will assume that the line averaged temperature is representative of the local electron temperature, i.e. $T_e = \langle T_e \rangle$, which is true if there are no strong temperature variations along the line of sight.

For generic values of electron density\cite{Nilson2006,Ross2012,Rosenberg2012} ranging between $10^{19}$ and $10^{20}$ cm$^{-3}$, and taking the measurements of $V_{\text{in}}$, $T_e$, and $Z$ discussed above, the experimental regime (with regards to heating) is characterized by

\begin{itemize}
    \item[1.] \emph{Weakly magnetized, flow-dominated plasmas} ($\beta_{\text{dyn}} > \beta_{\text{th}} \gg 1$). The hydromagnetic energy partition is defined by the thermal plasma-$\beta$ parameter $\beta_{\text{th}}$ (thermal-to-magnetic pressure ratio) and the dynamic plasma-$\beta$ parameter $\beta_{\text{dyn}}$ (ram-to-magnetic pressure ratio). This implies that most of the free energy density is in the bulk motion of the plasma.
    \item[2.] \emph{Free-streaming ions with collisional electron flows} ($\lambda_{ii}^{\text{inter}} \gg 2\delta \gg \lambda_{ee}^{\text{inter}}$). The interpenetration ion-ion mean free path $\lambda_{ii}^{\text{inter}}$ for typical plume conditions is of order of a millimeter, which is a several times larger than the reconnection layer width $2\delta$. This implies that the ions do not stagnate on the axis, thermalizing their energy. However, electron-ion mean free paths, $\lambda_{ei}^{\text{inter}}$ and $\lambda_{ei}^{\text{intra}}$, and electron-electron mean free paths for binary Coulomb collisions are sufficiently short such that they are all smaller or of the order of $2 \delta$. Here the labels "intra" and "inter" are used to denote mean-free-paths inside a single flow and between interpenetrating (or merging) flows, respectively.
\end{itemize}

\begin{table}
\caption{\label{tab:plasma parameters}Summary of relevant characteristic plasma parameters estimated from $\langle T_e \rangle$ at $t = 4$ ns. Ranges are shown by taking expected electron density minimal/maximal values of $10^{19}$ cm$^{-3}$ to $10^{20}$ cm$^{-3}$ using a dash symbol ($-$), respectively. We have taken an average charge stage $Z=6$ for all relevant calculations. Undefined values are presented with three dots (...).}
\begin{ruledtabular}
\begin{tabular}{lccc}
Parameter (units)                                                               &                      & Plumes            & Current sheet  \\ \hline
\begin{tabular}[c]{@{}l@{}}Electron \\ temperature (eV)\end{tabular}            & $T_e$                &$180$         & $280$    \\
Magnetic field (T)                                                              & $B$                  &$5$           & 0            \\
Inflow velocity (km/s)                                                          & $V_{\text{in}}$      &$230$         & 0            \\
\begin{tabular}[c]{@{}l@{}}Electron-ion\\ equilibration time (ns)\end{tabular}  &$\tau_{\text{eq}}^{e\backslash i}$&$3.5-0.4 $&$6-0.7$  \\
Radiative cooling time (ns)                                                     & $\tau_{\text{cool}}$ &$50-5$        & $90-9$       \\
\begin{tabular}[c]{@{}l@{}}Layer half-thickness\\ (mm)\end{tabular}             & $\delta$             &  ...         &   $0.15$   \\
\begin{tabular}[c]{@{}l@{}}Layer half-length\\ (mm)\end{tabular}                & $L_P/2$              &  ...         &   $> 2$         \\
Alfv\'en speed (km/s)                                                           & $V_A$                &$27-9$        & ...            \\
\begin{tabular}[c]{@{}l@{}}Ion-acoustic speed\\ (km/s)\end{tabular}             &$c_s$                 &$250$         & $315$    \\
\begin{tabular}[c]{@{}l@{}}Magnetic \\ diffusivity (cm$^2$/s)\end{tabular}      &$\eta$                &$1.6\times 10^4$        & $8\times 10^3$ \\

\begin{tabular}[c]{@{}l@{}}Ion-ion mean free path\\ ($\mu$m)\end{tabular}       &
\begin{tabular}[c]{@{}c@{}}$\lambda_{ii}^{\text{intra}}$\\$\lambda_{ii}^{\text{inter}}$\end{tabular}&
\begin{tabular}[c]{@{}c@{}}$0.4-0.04$\\$1500-150$\end{tabular} &
\begin{tabular}[c]{@{}c@{}}$1-0.1$\\...\end{tabular}       \\

\begin{tabular}[c]{@{}l@{}}Electron-electron mean\\ free path ($\mu$m)\end{tabular}       &
\begin{tabular}[c]{@{}c@{}}$\lambda_{ee}^{\text{intra}}$\\$\lambda_{ee}^{\text{inter}}$\end{tabular}&
\begin{tabular}[c]{@{}c@{}}$25-2.5$\\$1-0.1$\end{tabular} &
\begin{tabular}[c]{@{}c@{}}$65-6.5$\\...\end{tabular}       \\

\begin{tabular}[c]{@{}l@{}}Sonic Mach\\ number\end{tabular}                     & M$_\text{s}$         & $\lesssim 1$          & 0               \\
\begin{tabular}[c]{@{}l@{}}Alfv\'enic Mach\\ number\end{tabular}                & M$_{\text{A}}$       & $9-27$       & 0               \\
Thermal beta                                                                    & $\beta_{\text{th}}$  & $10-100$ & ...               \\
Dynamic beta                                                                    & $\beta_{\text{dyn}}$ & $50-500$ & ...                \\
Lundquist number                                                                & $S$                  & \multicolumn{2}{c}{ $60-20$ }                     
\end{tabular}
\end{ruledtabular}
\end{table}

These plasma parameters and others of interest for laser-driven magnetic reconnection in general are summarized in Table \ref{tab:plasma parameters}. We hypothesize that in the regime of interest, the four leading-order heating (or cooling) mechanisms are (i) plasma compression, (ii) electron-ion collisional heating, (iii) magnetic reconnection, and (iv) radiative cooling. Hence, the electron heating equation is given by
\begin{equation} \label{eq:Te evolution}
    \frac{3}{2}k_Bn_e\dot{T}_e = Q_{comp} + Q_{col}  + Q_{rec} + Q_{rad},
\end{equation}
where $Q_{\text{comp}} \equiv k_B \dot{n}_e T_e$ is the compression heating (or cooling) rate, $Q_{\text{col}} \equiv 4m_e\nu_{ei}n_eV_{\text{in}}^2$ is the electron-ion collisional drag heating (here $n_e$ is the inflow electron density), where
\begin{equation}
    \nu_{ei} = \frac{2.91\times10^{-6}}{\ln(\Lambda)} \frac{Z n_e[\text{cm}^{-3}]}{(T_e [\text{eV}])^{3/2}},
\end{equation}
 is the electron-ion collision rate\cite{Turnbull2023} (again, parameterized by the inflow electron density $n_e$), $\ln(\Lambda)$ is the Coulomb logarithm, $Q_{rec} = \mathbf{E}\cdot \mathbf{j}$ is the magnetic reconnection heating ($\mathbf{E}$ is the reconnected electric field and $\mathbf{j}$ the reconnected current density), and the radiative cooling power $Q_\text{rad} \propto \int j_\nu d\nu$. The characteristic cooling time for free-free emission is given by \cite{Rybicki2004}
 \begin{equation}
     \tau_{\text{cool}} = \frac{3n_e k_B T_e }{\int j_\nu d\nu}.
 \end{equation}
Taking $T_e = 250$ eV (approximately the temperature for both the plumes and the current sheet), the cooling time ranges from $\tau_{\text{cool}} \approx 8$ ns for $10^{20}$ cm$^{-3}$ density to $\tau_{\text{cool}} \approx 80$ ns for $10^{19}$ cm$^{-3}$ density. Since we observe the heating to act in $\sim 1$ ns timescales, we neglect this sink term in the equation (\ref{eq:Te evolution}). We provide more accurate values for plumes and reconnection layer separately in Table \ref{tab:plasma parameters}.

\begin{figure}
    \centering
    \includegraphics[width=7.5cm]{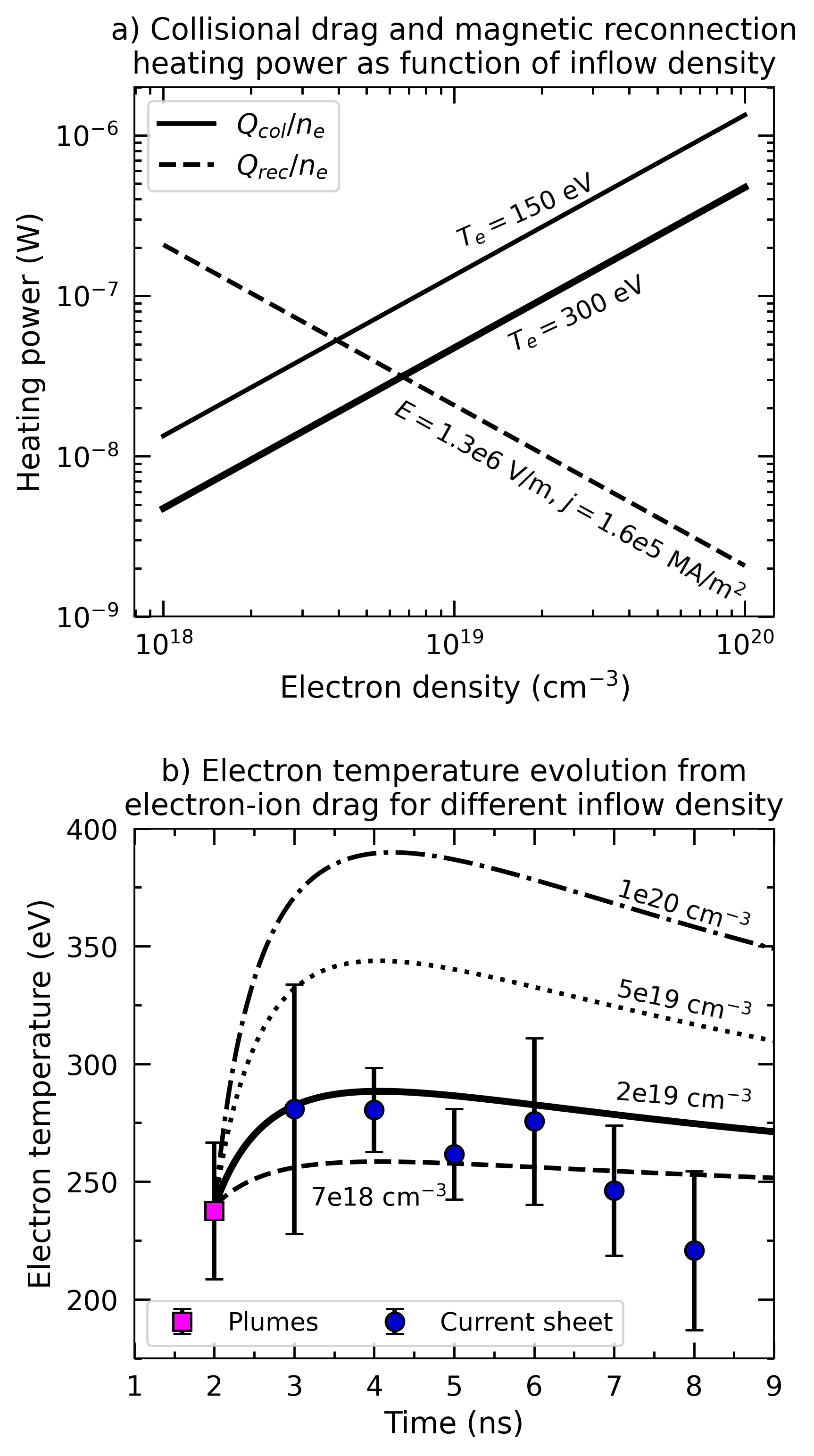}
    \caption{Calculations of magnetic reconnection and electron-ion drag heating. a) Comparison between magnetic reconnection $Q_{rec}$ heating and collisional heating $Q_{col}$, both normalized by density. We have assumed $V_{\text{in}}=230$ km/s. b) Electron temperature evolution as predicted by equations (\ref{eq:heating}) and (\ref{eq:density}) for different values of inflow density. Measured vales of $\bar{T}_e$ from Figure \ref{fig:plasma_evolution} are also presented.}
    \label{fig:heating}
\end{figure}

We now compare the magnetic reconnection against electron-ion drag heating effects. Given that the experiment is in the high-$\beta$ regime, we ask if the collisional friction between free streaming ions and electrons is efficient enough to dominate over reconnection heating. Figure \ref{fig:heating}a shows the heating power source terms (i.e. normalized by density) for two different values of $T_e$ as a function of density in the range of interest. We find that with increasing density, magnetic reconnection can provide decreasing heating power, and the contrary occurs for electron-ion collisional drag. Using electric field and current density values inferred from proton radiography\cite{Fox2020}, and considering a vertical length scale $\sim 1$ mm, the two sources of heating become comparable at relatively low densities of a few $\times 10^{18}$ cm$^{-3}$. Notice that $Q_{rec}$ would be overestimated if the field is advected away from the target along the $Z$ axis. This would increase the integration length-scale, reducing the magnetic field strength consistent with the measurement. If, on the contrary, the magnetic field remained confined close to the target's surface (supported by an extended magnetohydrodynamical effect such as the Nernst effect\cite{Walsh2020}), then we would potentially observe localized heating where magnetic reconnection is strong. We do not observe such features, so we conclude that this is unlikely. From this assumption, we find that collisional drag overcomes magnetic reconnection for any density $> 7 \times 10^{18}$ cm$^{-3}$ for the temperatures of both the plumes and current sheet. We conclude that conversion from magnetic to electron thermal energy is negligible compared to the frictional heating in the most likely conditions of interest (assuming the inflowing laser-drive plumes to have characteristic electron densities $n_e > 10^{19}$ cm$^{-3}$ at the NIF), and we therefore can drop $Q_{rec}$ from equation (\ref{eq:Te evolution}).

After dropping the reconnection heating source, we can explicitly calculate the current sheet's electron temperature evolution from electron-ion drag heating together with a compression term. The model by Ross et al.\cite{Ross2012,Ryutov2011} allows an analytical solution for the reduced equation (\ref{eq:Te evolution}), i.e. without radiative cooling nor magnetic reconnection terms. This model assumes two interpenetrating plasmas with free-streaming ions propagating head on, coming in contact at $t=t_0$, each with velocity (modulus) $V_\text{in}$, ion density $n_i$, and charge state $Z$. The energy exchange is mediated by electron-ion collisions and parameterized by the collision rate $\nu_{ei}$. Assuming quasi-neutrality in the interaction region, i.e. such that the overlapped electron density is $n_e = 2Zn_i$, the electron temperature evolution has an analytical solution given by
\begin{eqnarray} \label{eq:heating}
    \left[T_e(t)\right]^{5/2} &&= \left[T_e(t_0)\right]^{5/2} \\
    && + 56\times 10^{-39} Z^2 \ln(\Lambda) n_i(t)^{5/3} \int_{t_0}^t V_{\text{in}}(t')^2 n_i(t')^{-2/3}dt', \nonumber
\end{eqnarray}
where $\ln(\Lambda)$ is the Coulomb logarithm ($\approx 8$, in our experiments), and $t_0$ is the timing at which the interaction starts. The pre-factor in the equation is correct when $T_e$ is measured in keV, number densities in cm$^{-3}$, velocity in cm/s, and time in ns. We can simplify the calculation further by assuming a ballistic 3-dimensional expansion of the plasma plumes. Consider the plasma plumes to be rectangular boxes expanding with velocities $V_x$, $V_y$ and $V_z$ along the $XYZ-$directions, respectively, such that they are all constant but not necessarily equal. Assuming that once the laser drive is off, no additional plasma is created (or destroyed via recombination), the evolution of the density can be scaled from the non-conformal expansion of the plasma volume. As an initial condition, we assume the plumes to be identical with initial density $n_e(t_0) \equiv n_{e,0}$, such that the density evolution is given by
\begin{eqnarray}\label{eq:density}
    n_e(t) = n_{e,0}&&\times \left(\frac{x_0}{V_x(t-t_0)+x_0}\right)  \\
    &&\times \left(\frac{y_0}{V_y(t-t_0)+y_0}\right) \times \left(\frac{z_0}{V_z(t-t_0)+z_0}  \right), \nonumber
\end{eqnarray}
where $x_0$, $y_0$, $z_0$ are the plasma spatial dimensions at $t=t_0$. The product of the three terms in parenthesis yields the plasma volume at time $t$. Based on the observed plasma dynamics discussed in Section \ref{sec:results} A., we take $t_0=2$ ns as the earliest time when we have a clear characterization of the inflows, and as such $x_0 = z_0 = 1$ mm, $y_0=5$ mm, $V_x = V_y = V_{\text{in}} = 230$ km/s, and $V_z = 315$ km/s. We have checked that the result is not sensitive to assuming a different value of $V_y$, of which we do not have direct probing. By replacing equation (\ref{eq:density}) in equation (\ref{eq:heating}) and integrating numerically for three different values of $n_{e,0}$, the evolution of the electron temperature can be calculated. Figure \ref{fig:heating}b presents the result for different values of $n_{e,0}$ compared against the experimental data. We see best agreement with initial plume density $n_{e,0}=2\times 10^{19}$ cm$^{-3}$, with smaller values being insufficient to explain the observed electron heating. The dashed line shows that the initial expected heating in conditions where $Q_{col} \approx Q_{rec}$, further shows the small effect that magnetic reconnection has when trying to account for the data on timescales $\sim 1$ ns after initial plume-plume interaction. We note that on longer timescales, the first-order ballistic density evolution (equation (\ref{eq:density})) is no longer valid, as the flow velocity is not constant\cite{Ross2012}, hence we do not expect our electron-ion collision analysis to continue to be valid. It is possible that on those longer timescales, the combination of adiabatic expansion, heat transport, and radiative cooling lower the electron temperature of the current sheet, and account for the lower temperatures observed in the long-term plasma evolution.

The theoretical curves greatly overshoot above the data for larger densities, even in the expected range. It is possible that the initial density is larger than  $2\times 10^{19}$ cm$^{-3}$ and heat transported through the plumes can decrease the final electron temperature value. However, this calculation shows that in principle the plasma plumes have enough bulk kinetic energy and are collisional enough, such that electron-ion drag on its own can account for the observed temperature increase, and that in the same conditions, we expect magnetic reconnection to take a small role in the electron heating. It is interesting that, at late times $t>7$ ns, the inferred current sheet temperature is smaller than at $4$ ns. Again, this could be explained by a combination of adiabatic expansion and radiative cooling at late times. Although we argued above that radiative cooling is too slow to play a role in the initial $\sim 1$ ns heating timescales, it could still contribute to the plasma cooling on scales $\sim 10$ ns. Future experiments with detailed (both spatially and temporally resolved) measurements could be used to test this hypothesis.

\section{\label{sec:conclusions}Conclusions}
In summary, we have reported a spatially- and temporally-resolved characterization of electron temperature in magnetic reconnection experiments at the NIF using filtered X-ray self-emission images. We have developed and used a statistical approach which allows accounting for both systematic and random errors introduced by the MCP gain and plasma evolution, amongst others. Our main conclusions are:
\begin{itemize}
    \item[1. ]We have demonstrated a method to infer line integrated, 2-D maps of electron temperature using soft X-ray plasma self-emission at the National Ignition Facility. Using redundant measurements, it is possible to identify and marginalize over systematic errors. The main sources of uncertainty stem from low photon counts after filtration and the MCP gain. These are visually apparent on the maps as speckled patterns on the temperature maps that correlate with locally higher photon counts, but do not correspond to spatial modulations of the plasma temperature.  
    \item[2. ]After formation, the current sheet significantly heats up compared to the inflowing plumes, and then remains at an approximate constant temperature $280 \pm 20$ eV (at $t=4$ ns), with a slight decline, reaching $220 \pm 30$ eV at 8 ns. In contrast, the plumes have an initial electron temperature of $240 \pm 20$ eV and show weak cooling with a final temperature of $180 \pm 30$ eV at $t=4$ ns, after which the plume self-emission intensity is too low to be detectable in the used configuration. Both regions show a slight temperature decrease later in time.
    \item[3. ] For a range of expected densities we show that magnetic reconnection is an insufficient heat source to account for the temperature separation between plumes and reconnection layer. Using a semi-analytic model we argue that, for the same range of parameters, electron-ion drag heating is a more likely candidate to explain our results.  
\end{itemize}

Future experiments could use imaging optical Thomson scattering measurements across the plume-layer structure to directly probe $V_{\text{in}}$, $T_e$, and $T_i$ in the plumes and current sheet. A free space configuration of the Thomson scattering probe would allow the measurement of the density and heat transported through the plumes from the reconnection layer through space. In addition, we suggest studying the effect of the current sheet temperature on plasmoid formation\cite{Fox2020}. Pulsed-power experiments in the semi-collisional plasmoid unstable regime have found anomalous ion heating\cite{Hare2017a}, which we cannot measure using GXD diagnostics but are accessible through Thomson scattering, and therefore would be an interesting next step for investigation.

\begin{acknowledgments}
 We thank the NIF Discovery Science Program for facilitating these experiments. Support for these experiments was provided by the U.S. DOE, Office of Fusion Energy Sciences, under FWP SW1626 FES. This work was performed under the auspices of the U.S. DOE by Lawrence Livermore National Laboratory under Contract No. DE-AC52-07NA27344 and under DOE Award Nos. DE- SC-0016249, DE-NA-0003856, DE-NA0004144, and Field Work Proposal No. 4507 under DOE Contract No. DE-AC02-09CH11466. We thank Jack Halliday for suggesting we assess the effect of the carbon recombination feature in the analysis, and Frances Kraus for running FLYCHK simulations.
\end{acknowledgments}

\section*{Data Availability Statement}

Data were generated at a central, large scale facility. Raw data were generated at the National Ignition Facility. Derived data supporting the findings of this study are available from the corresponding author upon reasonable request.

\appendix

\section{\label{app:HIPPIE} The \hip  code}

\subsection{Summary}
The X-ray images were analysed using the \texttt{MATLAB}-based pinHole Imaging PiPlInE (\texttt{HIPPIE}) code. The code imports the hdf5 files from an experimental CCD image together with a background flat-field dataset as calibration. The flat-field image was obtained using the same camera and operational settings. Alternatively, the code allows to generate synthetic 2-D flat fields using a polynomial profile as droop function and arbitrary gain for the strips. The flat-fielding can be skipped by setting the droop function to $1$.

\begin{figure}
    \centering
    \includegraphics[width=8cm]{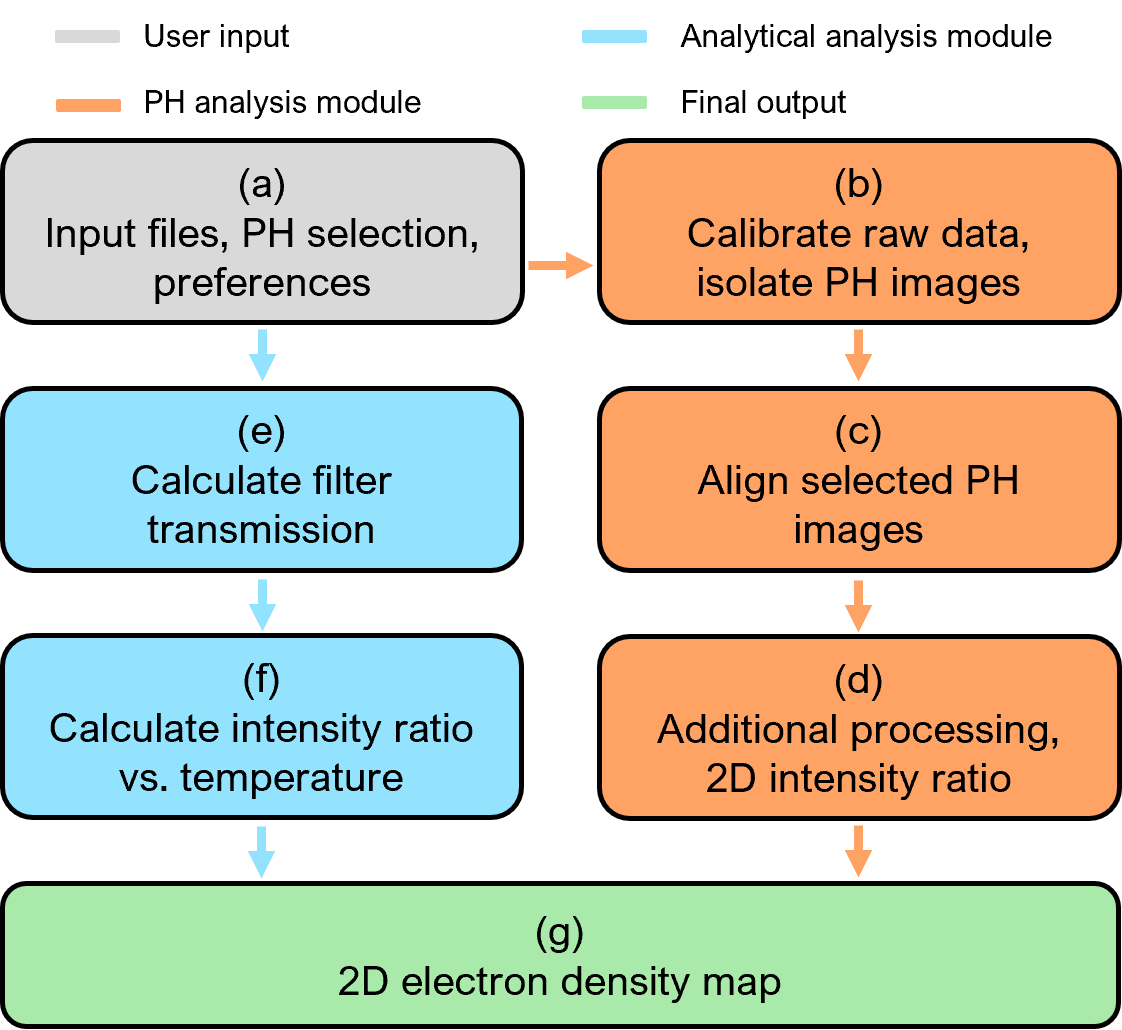}
    \caption{High-level flow chart overview of \hip.  User inputs are depicted in the grey box (a). Preferences include smoothing and binning settings. The orange processes (a-c) represent the pinhole analysis module, the light blue boxes (d,e) denote the analytical module, and the green box (f) denotes the the 2D electron temperature map as the final output.}
    \label{fig:pipelineFlow}
\end{figure}

The pipeline schematic is presented in Figure \ref{fig:pipelineFlow}. \hip is split into two concomitant parts: a pinhole analysis module (panels b-d) in which the x-ray data is processed to produce the left-hand side of equation (\ref{eq:ratio}), and an analytical analysis module (panels e-f) that calculates the right-hand side.

The pinhole (PH) analysis part is organized as follows. The pipeline retrieves the CCD image and subtracts the dark current using a non-illuminated region of the CCD. Afterwards, the plasma image is divided by the flat-field. The user specifies the pinhole position of the images of interest together with geometric constraints. The user defines the images to be compared and the code isolates the field of view of the requested pinhole images. Then, the code aligns the lower intensity image with respect to the higher intensity image. After that, it subtracts the X-ray background on both images and takes the ratio of intensity.


The pinhole selection information specifies which two pinhole images are compared, and the alignment preferences specify a) whether a user would like \hip to operate in automated or manual alignment mode and b) which feature to align with respect to (e.g., left plume, current sheet, or right plume). This is useful for analyzing different stages of the plasma evolution (e.g. before or after current sheet formation) and for aligning images with poor photon statistics on given regions of the images. Finally, the smoothing and binning preferences specify a) how much to smooth the images \textit{prior to alignment} and b) the binning method and bin size by which to bin the image after alignment. Here, a) allows the user to lightly smooth each individual pinhole image to aid in alignment and b) serves to give the user the ability to downscale the 2D intensity ratio, i.e. left-hand-side of equation (\ref{eq:ratio}). 

\subsection{X-ray Image Calibration, Alignment, and Ratio Construction}\label{expAnalysis}

With the initial inputs from Section \ref{analysisSection} specified, \hip begins with step (b) as defined by Figure \ref{fig:pipelineFlow}. The data file path information is used to read in the hdf5 files provided by NIF for each shot and its corresponding calibration image. For both the shot and its calibration shot, an initial background subtraction takes place that subtracts off a pre-shot image with ambient signal from the shot image with x-ray data. Both images are floored at 0 after this initial background subtraction. We define a flat field image as the calibration image divided by the mean of the calibration image. Dividing the background-subtracted shot image by the flat field image produces the "raw" image from which we work. The user-specified pinhole selection information is used to pick out which two pinholes to use for the analysis. Each pinhole is isolated from the raw image such that the local maxima of the user-specified feature of interest (left plume, current sheet, or right plume) is in roughly the same place within a 1050 x 1050 image for both pinholes. To aid this process, we apply a light 2D Gaussian filter to each pinhole so as to make local maxima identification more accurate. This alignment via the maxima only serves to get the pinholes \textit{roughly} aligned - i.e., globally in the right place - which reduces the work done by the manual and automated alignment algorithms further down the pipline. 

With the pinholes now isolated, \hip moves into its alignment phase (step (c) in Figure \ref{fig:pipelineFlow}). Depending on the alignment choice specified by the user, \hip will either enter a manual alignment mode or one that is automated. In both cases, the high signal pinhole is held fixed while the low signal pinhole is moved around until alignment is achieved. If the user wishes to align the pinholes manually, \hip will display vertical and horizontal lineouts through the local maxima of the feature of interest, as well as the full pinhole image, for both pinholes and prompt the user for the number of pixels by which to shift the low-signal pinhole. This prompt will continue until the user is satisfied with the alignment, at which point they may exit the loop and continue on to step (d). Should the user wish to automate the pinhole alignment, \hip will iteratively approach an aligned configuration by first aligning the pinholes at the target edge (vertical alignment) and then aligning the pinholes horizontally at the inflection point of both horizontal lineouts. This process will continue until either the algorithm finds no further shifting is necessary in both directions or the algorithm gets stuck. In the case of the latter, \hip will notify the user that the automated alignment has failed and terminate all processes so that the user may \textit{manually} align the pinholes instead.

\begin{figure}
    \centering
    \includegraphics[width=8.5cm]{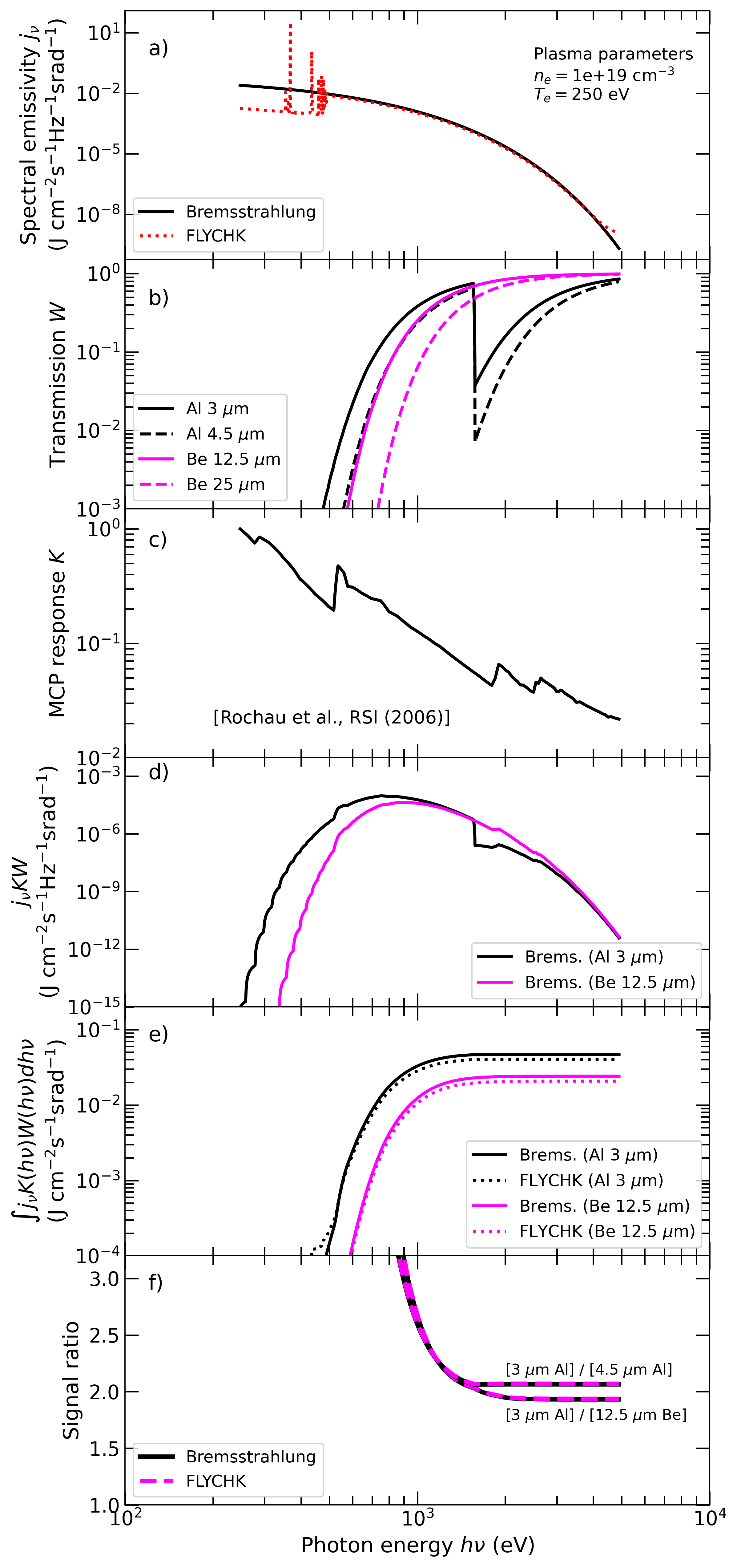}
    \caption{Emissivity, transmission, and detector calculations in the regime of interest. a) Plasma spectral emissivity calculated for representative plasma parameters. Red dotted line shows FLYCHK\cite{Chung2005} calculations which contain low-energy free-bound electron emission. Solid line shows the bremsstrahlung component in the same conditions. b) Filter transmissions\cite{Henke1993,Oertel2006} using the 'checkerboard' pattern configuration. c) MCP spectral response\cite{Rochau2006}. d) Spectral emissivity modulated by two example filter transmissions and MCP response. e) Integrated broadband intensity from panel (d) for FLYCHK (dotted) emissivity and bremsstrahlung component (solid). f) Signal ratio between filtered intensities. Bremsstrahlung calculations are shown in solid black line, whereas FLYCHK calculation in dashed magenta lines. Filter responses used for each curve are annotated.}
    \label{fig:filter_transmission}
\end{figure}

Once out of the alignment phase, \hip begins additional processing and construction of the intensity ratio, step (d) in Figure \ref{fig:pipelineFlow}. The next step is removing local background signal background subtraction and flooring. This additional background subtraction, different than that which occurs in step (b), is done by taking the mean value of $\sim 10$ pixels of ambient signal near the edges of the target, and subtracting that from the pinhole image. Once the background subtraction is done, the pinhole image is then floored to a user-specified value by setting anything beneath the floor limit to a NaN value. The ratio of the two pinholes is calculated by dividing the low-signal image by the high-signal image to produce the desired 2D intensity ratio. This 2D intensity ratio is then binned, if the user has specified to do so. The binning algorithm utilized by \hip was developed in-house for this diagnostic technique in particular. The user is allowed to specify a bin of arbitrary size (a user-specified bin-size of $1 \times 1$ bypasses the binning algorithm entirely) as well as how they would like to bin a given image (e.g., mean, median, geometric mean); it is simple to add another method of binning should another be needed. Closing off step (d) of \hip, and the entire pinhole analysis module, is the removal of the vacuum region (i.e., the region of the image above the target) simply for aesthetic reasons. This does not affect the quantitative analysis in any way. The final data product exported by the pinhole analysis module is the 2D intensity ratio of the two pinholes specified by the user.\\ 

We now discuss the calculation of the Full-Width at Half-Maximum (FWHM) in object-plane units of the 2D gaussian filter. The CCD camera used throughout all of these experiments consists of $4096$ pixels $\times$ $4096$ pixels with a 9 $\mu$m pixel size. A given shot configuration produces a magnification of $M$, then we may calculate the spatial resolution in the object plane is $36/4200 M$ mm/pixel.

In the analysis, we chose the smoothing to a $\sigma_{\text{pixels}} = 10$ pixels. Thus, we can convert this into physical units within the object plane as $\sigma_{\text{mm}} = 3 \sigma_{\text{pixels}}/350M$ mm. Finally, the FWHM is given simply by FWHM $=2 \sqrt{2\ln(2)} \sigma_{\text{mm}} \approx 0.02M$ mm. The experiments were fielded with a magnification $M=1.75$, and therefore the 2D gaussian filter has a FWHM $\approx 115$ $\mu$m.

\subsection{Analytical Calculation of Expected X-Ray Photon Yield}

\begin{figure*}
    \centering
    \includegraphics[width=17cm]{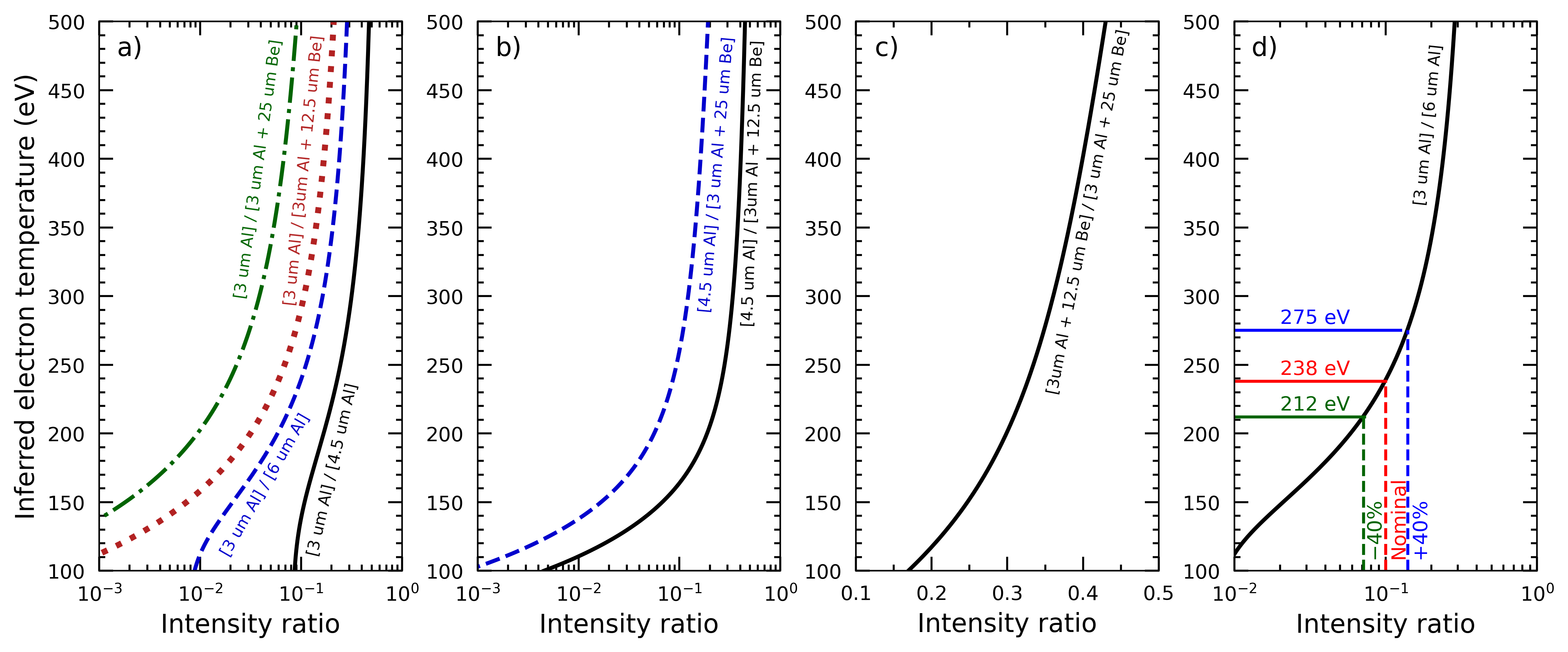}
    \caption{Tabulated values of electron temperature as a function of intensity ratio for filters presented in Table \ref{tab:filters}. (a) Curves calculated for $3$ $\mu$m Al as softest filter. (b) Curves calculated for $4.5$ $\mu$m Al as softest filter. (c) Curves calculated for $3$ $\mu$m Al $+$ $12.5$ $\mu$m Be as softest filter. (d) Zoom in to curve of interest for $t=2$ ns dataset. A nominal value of $0.1$ is shown as representative of the intensity ratio for pairs of images taken at the location along the MCP strip. The strips can have differences of up to $\pm 40\%$ in gain at the same bias voltage, which would lead to inferred electron temperatures in the range shown.}
    \label{fig:intensity ratio temperature}
\end{figure*}

The analysis method relies on the fact that the plasma emissivity is strongly dominated by bremsstrahlung continuum emission. Although carbon becomes fully ionized in the regime of interest, the spectrum exhibits free-bound recombination features at photon energies $\lesssim 500$ eV. Figure \ref{fig:filter_transmission}a shows a comparison of the spectral emissivity using the atomic kinetic code FLYCHK and the analytic bremsstrahlung emission. We show that the high-pass filters used in the experiments with transmission coefficients presented in Figure \ref{fig:filter_transmission}b are strong enough to remove the bright recombination features. The MCP spectral response is shown in Figure \ref{fig:filter_transmission}c.

\begin{figure*}
    \centering
    \includegraphics[width=16cm]{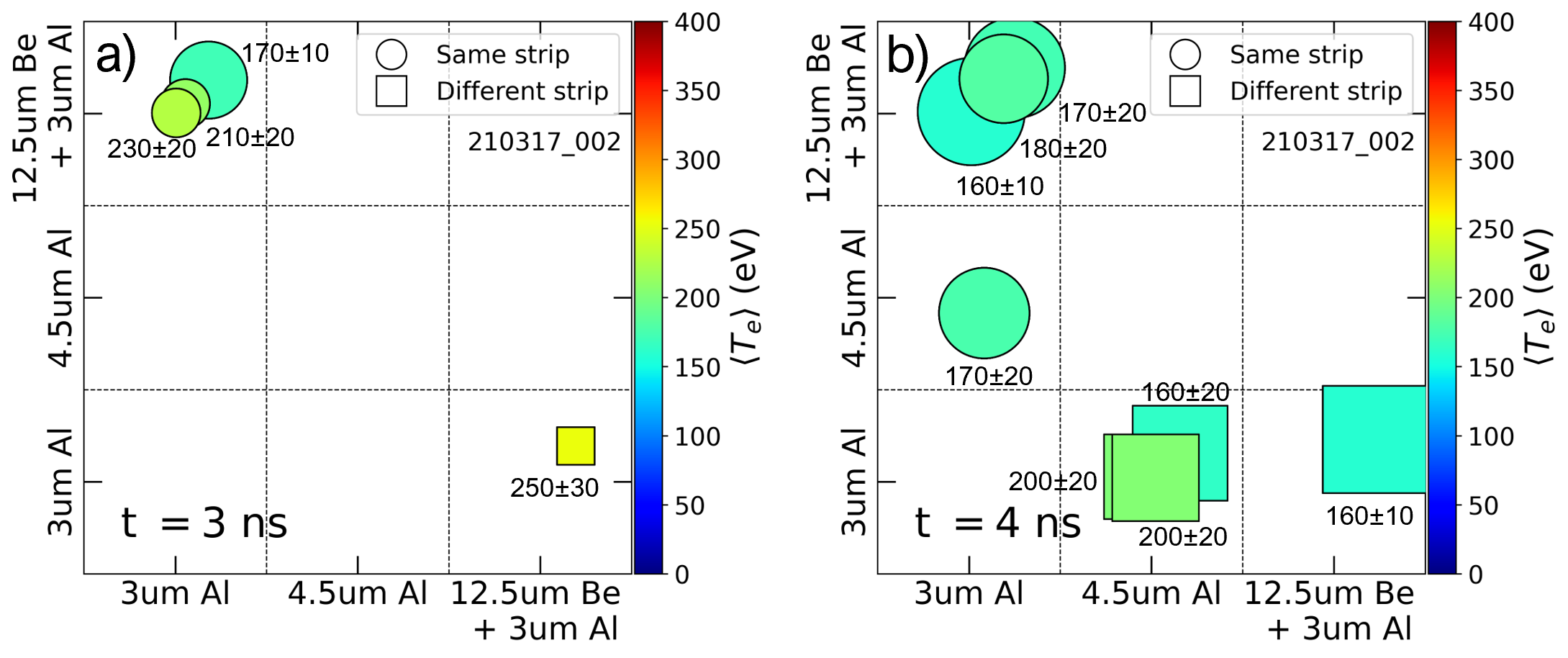}
    \caption{Spatially-averaged electron temperature diagrams of the plumes' inferred from different pinhole/filter.}
    \label{fig:plume_temp_summary}
\end{figure*}

Figure \ref{fig:filter_transmission}d shows that the spectral emissivity convolved with the foil transmission, acting as a high-pass filter, and MCP response, acting as a low-pass filter, result in a band-pass filter strongly peaked at photon energies around $h\nu=1$ keV. Panel (e) shows the integration with photon energy, which shows very good convergence of the total convolved emission of the kinetic code and bremsstrahlung emission. Most importantly, small disagreement ($\sim 10\%$ relative change) is the same for both filters. We therefore conclude that the low-energy recombination features do not significantly contribute to the overall detected signal nor temperature inference. Most importantly, the contribution to the signal ($\propto \int j_\nu K(\nu) W(\nu) d\nu)$) is independent from the filter used. Panel (f) shows the signal ratio for two types of filters, with 3 $\mu$m Al as soft filter, and either 4.5 $\mu$m Al or 12.5 $\mu$m Be as hard filter. The ratio does not change due to the recombination spectral features and therefore the diagnostic is insensitive to their contribution.

\begin{figure*}
    \centering
    \includegraphics[width=16cm]{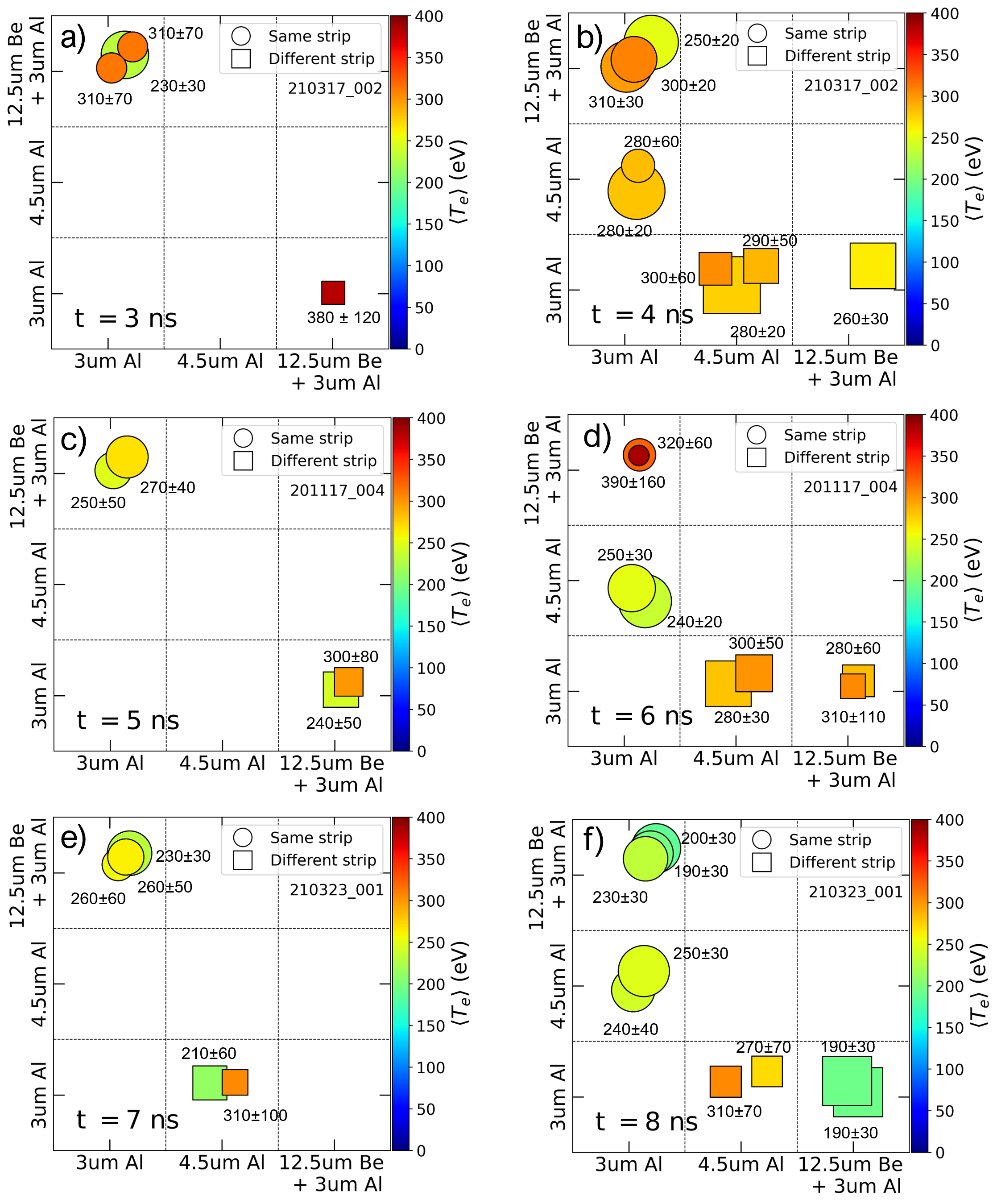}
    \caption{Spatially-averaged electron temperature diagrams of the current sheet inferred from different pinhole/filter.}
    \label{fig:temp_summary}
\end{figure*}

Steps (e-f) of \hip entail the calculation of the right-hand side of equation \ref{eq:ratio}. In the following stages, \hip interfaces with user-supplied files: 1) a spreadsheet containing the filter material names and thicknesses for the pinholes analyzed in steps (b-d), and 2) filter transmission \cite{Henke1993}. The filter transmission is presented in Figure \ref{fig:filter_transmission}.

Step (e) is dedicated to the calculation of $W_{1}(\nu)$ and $W_{2}(\nu)$. \hip utilizes the aforementioned spreadsheet to extract the filter materials and thicknesses for the pinholes that the user has chosen to analyze. The code constructs the appropriate file name for (and imports) the transmission coefficient data for each material in use. These dataset\cite{Henke1993} inputs are at a standard thickness of $10$ $\mu$m, which \hip will then re-scale appropriately according to
\begin{equation}
    W(d) = W_{0} ^ {\frac{d}{d_{0}}},
\end{equation}
where $W_{0}$ is the transmission coefficient data for a given material at a standard thickness of $d_{0} = 10$ $\mu$m, and $W$ is the re-scaled transmission coefficient data for the same material at the actual filter thickness $d$ used in the experiment. After re-scaling, and if a pinhole was filtered by two stacked filters, $W_A$ and $W_B$, the two transmission coefficient data sets are combined as
\begin{equation}
    W_{\text{total}} = W_A W_B,
\end{equation}
thus for stacked filters, $W_{\text{total}}$ goes into the integral in equation (\ref{eq:ratio}). Step (f) calculates these integrals and constructs the expected ratio of X-ray photons per pinhole. To this end, $j_{\nu}$ (equation \ref{eq:emissivity}) is a $0^{th}$-order modified Bessel function of the second kind, $W(\nu)$ has already been calculated in the previous step, and the spectral response function $K(\nu)$ is imported from the work of Rochau \textit{et al} \cite{Rochau2006}. After integration and division, the final data product of step (f), and the entire analytical analysis module, is the expected ratio of X-ray photons emitted per pinhole, which is solely a function of $T_{e}$.

\subsection{Temperature Calculation}

This final stage of \hip (step (g) in Figure \ref{fig:pipelineFlow}) is the construction of the 2D temperature map from the data products produced by both the pinhole and analytical analysis modules. \hip uses the filter and MCP datasets to forward-model equation (\ref{eq:ratio}) and creates look-up tables relating the intensity with the electron temperature. In the conditions and configuration of interest, the tabulated values are presented in Figure \ref{fig:intensity ratio temperature}. The code then iterates through the entire 2D intensity ratio image produced from the experimental data and creates the temperature map.



\section{Filter-filter diagrams}
The filter-filter diagrams corresponding to the plumes and current sheet for all 'checkerboard' datasets are presented in Figures \ref{fig:plume_temp_summary} and \ref{fig:temp_summary}. These were used to flag systematic discrepancies when using images taken on different portions of the detector (either along a single strip, across different strips, or both). We note that the dynamic range of strips 1 and 2 is narrower than strips 3 and 4, and therefore the latter yielded more $\langle T_e \rangle$ maps. This does not affect our conclusions.



\nocite{*}
\bibliography{aipsamp}

\end{document}